\documentclass[11pt]{article}
\usepackage{vmargin}
\setpapersize{USletter}           
\setmargrb{1.4in}{0.5in}{1.4in}{1in}

\usepackage{xcolor}
\usepackage{latexsym}
\usepackage{graphicx}
\usepackage{amssymb, amsmath, bbold}
\usepackage{epstopdf}
\usepackage{float}
\usepackage{hyperref}
\usepackage{mathtools}
\usepackage{ulem}
\usepackage{cancel}
\usepackage{lineno}

\definecolor{green}{rgb}{0,0.8,0}
\definecolor{red}{rgb}{1,0,0}
\definecolor{blue}{rgb}{0,0,1}
\definecolor{purple}{rgb}{0.7,0,0.7}

\def\be{\begin{equation}}
\def\ee{\end{equation}}
\def\ba{\begin{array}}
\def\ea{\end{array}}
\def\noi{\noindent}
\def\sdot{\mbox{${\dot{s}}$}}
\def\veps{\mbox{$\varepsilon$}}

\begin{document}



\title{On the dissipation at a shock wave in an elastic bar}

\author{Prashant K. Purohit$^1$ and Rohan Abeyaratne$^{2}$\\[2ex]
$^1$Department of Mechanical Engineering and Applied Mechanics, \\
University of Pennsylvania,\\
Philadelphia, PA, 19104, USA\\
purohit@seas.upenn.edu\\[2ex]
$^2$Department of Mechanical Engineering\\
Massachusetts Institute of Technology\\
Cambridge, MA 02139, USA\\
rohan@mit.edu\\[2ex]
}

\renewcommand{\thefootnote}{\arabic{footnote}}
\setcounter{footnote}{0}

\maketitle

\begin{abstract}

This paper aims to quantitatively relate the  energy dissipated at a shock wave in a nonlinearly elastic bar to the energy in the oscillations in two related dissipationless, dispersive systems.  In contrast to a phase boundary, there is no kinetic relation associated with a shock wave. Three one-dimensional dynamic impact problems are studied:  Problem 1 concerns a nonlinearly elastic bar, Problem 2 a discrete chain of particles, and Problem 3 a continuum with a strain gradient term in the constitutive relation.  In the impact problem considered, the free boundary of each initially quiescent body is subjected to a sudden velocity that is then held constant for all subsequent time.  There is energy dissipation at the shock in Problem 1 but Problems 2 and 3 are conservative. Problem 1 is solved analytically, Problem 2 numerically and an approximate solution to Problem 3 is constructed using modulation theory. The rate of increase of the oscillatory energy in Problems 2 and 3 are calculated and compared with  the dissipation rate at the shock in Problem 1. The results indicate that the former is a good measure of the latter.

\end{abstract}

\noi Keywords: Shock wave, dissipation, dispersion, elastic bar, discrete chain, dispersive shock wave, oscillations



\section{Introduction.}

Dissipation in an elastic body sounds like a contradiction since we often think of elastic as being synonymous with dissipationless. However, if an elastic body, even a hyperelastic body, involves a moving singularity such as a shock wave\footnote{other examples include a propagating crack or dislocation,}, there is a loss of energy at the singularity.  This is usually attributed to a deficiency in the elastic model, at least when it comes to describing such a feature.  This leads to ``regularization'' of the model which entails accounting for other physical effects. For example a dissipative regularization involves adding, say, a viscous term to the elastic constitutive relation, and this causes the sharp elastic shock fronts to turn into narrow zones in which the fields vary continuously (but rapidly).  However, if one wants to examine the dissipation in the elastic body as we do, supplementing the model with additional sources of dissipation is not the best way to proceed.  Therefore we shall not pursue  dissipative regularizations of the elasticity problem.

On the other hand a dissipationless, dispersive regularization would involve, say, adding a conservative strain-gradient term to the elastic constitutive equation. The shock wave in the elastic body is now replaced by a dispersive wave packet. It is usually claimed that the energy in the oscillations of the wave packet correspond to the energy dissipated at the shock wave in the elastic body. While this is certainly plausible and likely, we have not found a quantitative demonstration of this fact in the literature, and that is the focus of this paper.

We consider three closely related problems.  Problem 1 concerns a semi-infinite nonlinearly elastic bar.  The bar is initially stress free and at rest.  At time $t=0^+$ its free boundary is given a speed $V$ which is held constant from then on.  A shock wave emerges from the loading surface $x=0$ and propagates into the quiescent material  at a constant speed. The strain $\gamma^-$ and particle speed $v^-=-V$ behind the shock are constant. The problem can be readily solved analytically and, in particular, the dissipation rate calculated explicitly.

Problem 2 is a discrete counterpart of Problem 1.   It involves a semi-infinite row of identical particles 
with each particle interacting with its nearest neighbors through identical nonlinearly elastic springs.  The force-displacement relation of a spring is related to the stress-strain relation of the continuum. This system is dissipationless.  The spacing between the particles introduces a length scale into the problem and the sudden loading causes a dispersive wave packet to propagate into the quiescent material. We solve this problem numerically and various features of the solution are determined.

 Finally Problem 3 again concerns a continuum.  It is like Problem 1 except that the constitutive relation is augmented with a linear strain-gradient term.  This higher gradient term introduces a length scale.  This too is a dispersive conservative system. We construct an approximate solution to this problem using Whitham's theory of modulated waves, \cite{Whitham1965a, Whitham1965b, Whitham1970}.

It is important to emphasize that since we want the shock wave to be the {\it only} source of dissipation in Problem 1,  the nonlinear stress-strain relation characterizing the material is taken to be monotonic and convex. This prevents the occurrence, for example, of phase transitions which have their own dissipation. Having other sources of dissipation would only muddy the central question we want to study.  Some references to the  literature on discrete and continuous systems undergoing phase transitions will be given below.

Similarly, we emphasize that we are concerned entirely with the three aforementioned {\it mechanical} problems.  In understanding the relation between their energetics, their ``energy budgets'', we do not wish to bring in either temperature/thermodynamics or statistical mechanics.  We want to answer our question within the framework of classical mechanics.

After the free end of the system has been subjected to the impact speed $V$, there is a uniform state behind the propagating shock wave (in Problem 1) or behind the dispersive wave packets (in Problems 2 and 3). The particle speed in this region is $v^-=-V$ and the value of the strain, $\gamma^-$, is determined by solving the relevant equations (which are different in the three problems).  Our first observation was that this strain was essentially the same in all three problems (though not strictly identical); see Figures \ref{Fig-1.pdf} and \ref{Fig-V-gammaminus} noting that one of the curves in the latter figure has been shifted for clarity.  The rate of dissipation $\Bbb D$ in Problem 1 can be readily calculated in terms of $V$.  In Problems 2 and 3 we let $\gamma_{\rm osc}$ and $v_{\rm osc}$ denote the oscillatory part of the strain and particle speed (in two different senses that will be defined in Section \ref{sec-3}). We then calculate the energy in the system based on $\gamma_{\rm osc}$ and $v_{\rm osc}$ which we refer to as the ``excess'' or ``oscillatory'' energy in the system.  The rate of increase of the oscillatory energy is then compared with $\Bbb D$.  In Problem 2 this is done numerically.   In Problem 3 the wave packet involves a slowly varying amplitude and a fast oscillation\footnote{This is also true in Problem 2 but since the solution is obtained numerically, we didn't find it necessary to average the solution.} , and so we average the oscillatory energy over the fast oscillations.  The results are shown in Figures \ref{Fig-DissProb2.pdf} and \ref{Fig-DissipationRate.pdf}.  They indicate that the rate of increase of oscillatory energy in Problems 2 and 3 is a good measure of the dissipation rate in Problem 1, though our results do not constitute a rigorous proof of this claim since are solutions our numerical (Problem 2) or approximate (Problem 3).

When a conservative dispersive system involves a propagating ``defect'' such as a dislocation or phase boundary, the energy radiated by the waves traveling away from the defect can be identified with a kinetic relation, and therefore with effective dissipation. This has been noted and explored in, for example, a Frenkel-Kontorova dislocation by Atkinson and Cabrera \cite{AC1965}, in phase transformations by Kresse-Truskinovsky-Vainchtein, \cite{Kresse2003, Trusk2005a}, and for a Peierls dislocation in two-dimensions by Sharma \cite{Sharma2005}. In our context, there is no kinetic relation associated with the motion of a shock wave, and indeed our choice of problem was dictated by this.

There is a rich literature on the dynamics of one-dimensional lattices. A few of these papers include: the celebrated Fermi-Pasta-Ulam-Tsingou (FPUT) problem where the authors investigated the transfer of energy between modes in a one-dimensional chain of particles, \cite{FPUT}; the closed form solution to a dynamic problem for a harmonic chain, Synge \cite{Synge1973} and Chin \cite{Chin1975}; the motion of a Frenkel-Kontorova dislocation, e.g.
Atkinson and Cabrera \cite{AC1965}; the dynamics of phase transitions, e.g. Kresse and Truskinovsky, \cite{Kresse2003}, Truskinovsky and Vainchtein  \cite{Trusk2005a}, Puglisi and Truskinovsky \cite{Trusk1990} and Purohit and Bhattacharya \cite{PKPKB2003}; the dispersive evolution of pulses in a lattice, e.g. Giannoulis, Herrmann and Mielke \cite{GM-2005};  the derivation by Aubry and Proville \cite{Aubry2009}  of Rankine-Hugoniot type jump conditions for a discrete damped nonlinear lattice; and so on.

The rigorous transition from a discrete model to a continuous one is subtle, e.g. see Giannoulis, Herrmann and Mielke \cite{gia-mielke-2006}.  Depending on the specific class of ``microscopic motions'' considered, the same discrete model will yield different continuum models, e.g. the KdV equation \cite{FrieseckePego}, the Schr\"{o}dinger equation \cite{gia-mielke-2004}, and of course the equations of classical elasticity.

There is likewise a vast literature on the dynamics of dispersive continuous systems. A subset of these are concerned with the motion of ``dispersive shock waves'' (DSWs) -- the dispersive non-dissipative counterpart of a shock wave. This body of work stemmed from the seminal ideas of Whitham \cite{Whitham1965a, Whitham1965b, Whitham1970} that have since been advanced by other researchers and used to study DSWs in compressible fluids, Bose-Einstein condensates, shallow water etc.; e.g. see the review article by El and Hoefer \cite{elhoefer}, the dissertation by Nguyen \cite{Nguyen}, the book by Kamchatnov \cite{Kamchatnov2000} and the references therein. Rigorous analyses include the work of 
Lax and Levermore \cite{LaxLev1, LaxLev2, LaxLev3}, Gurevich and  Pitaevskii \cite{GurevichPita-a, GurevichPita-b} and Venakides \cite{Venakides1985}. In a recent paper Gavrilyuk et al. \cite{Trusk2020} explore shock-like fronts in dispersive systems. The motion of DSWs in discrete particle chains have been explored by, e.g., Dreyer and Hermann \cite{Dreyer-2008}, and the equations of continuum thermomechanics (except the entropy {\it inequality}) have been derived from a discrete particle chain using modulation theory  by Dreyer, Hermann and Mielke \cite{DreyerHerrmannMielke2005}.

The basic idea underlying Whitham's modulation theory is that the dispersive wave packets of interest involve both slow and fast scales, e.g. an amplitude that varies on a slow scale and oscillations that occur on a fast scale. The approach to generating such solutions is to first find an (exact) periodic traveling wave involving some free parameters, and to then allow the parameters to vary slowly. These slowly varying quantities are determined from the modulation equations.  Our formulation of Problem 3 is Lagrangian (as is often the case in solid mechanics) and the governing equation for strain is decoupled from the second equation that involves both strain and particle speed. As a result, the traveling wave for strain involves three parameters, and that for the particle speed involves one more. This allows us to deal with the slow modulation of the parameters in two steps: first working on the parameters in the strain, and thereafter the one in the particle speed.  If we were to write the governing equations in Eulerian form, they would have the same structure as the equations governing the flow of a compressible fluid where Eulerian formulations are  customary (Section S4 of electronic supplemental material). DSWs in the latter system of equations have been studied in the literature, e.g. see Section IV-D of \cite{hoefer-2006}, but because of the coupling of the Eulerian equations, the modulation of the four parameters have to be dealt with simultaneously.

The organization of this paper is straightforward. 
Section \ref{sec-1} is devoted to Problem 1 (the elastic bar), Section \ref{sec-2} to Problem 2 (the discrete particle chain), and Section \ref{sec-3} to Problem 3 (the dispersive continuum model with strain-gradient effects).  We derive an explicit relation \eqref{eq-20210623-1}  between the states $\gamma^-, v^-$ behind the DSW and the state $\gamma^+, v^+$ ahead of it.  It is the counterpart of a Rankine-Hugoniot jump condition at a shock and the similar integral relation at a fan;  see also Gavrilyuk et al. \cite{Trusk2020}. For both Problems 2 and 3 we calculate the rate of increase of the oscillatory energy ($D$), and compare it with  the dissipation rate at the shock in Problem 1 ($\Bbb D$). The results shown in Figures \ref{Fig-DissProb2.pdf} and \ref{Fig-DissipationRate.pdf} suggest that $D$ is a good measure of $\Bbb D$.  Some additional details can be found in the supplemental material.

Finally we note that when we plot strain, impact speed, dissipation etc. we will scale them as $\gamma/\beta_0^2, V/(c_0 \beta_0^2), {\Bbb D}/(\mu c_0 \beta_0^4)$ etc. where $\mu, c_0$ and $\beta_0$ are three parameters in the problem. In that way, our results don't depend on the particular values of these parameters.


\section{Impact problem for a one-dimensional elastic continuum.} \label{sec-1}

In this section we consider the motion of a semi-infinite, one-dimensional, elastic bar.  A generic particle is identified by its location $x \geq 0$ in a stress-free reference configuration.  It is located at $y(x, t)$ at time $t$. The strain $\gamma(x, t)$, particle speed $v(x, t)$ and stress $\sigma(x, t)$ satisfy the equations
\be
\label{eq-20210512-1}
\gamma = y_x - 1, \quad v = y_t,  \quad  \sigma_x = \rho v_t, \qquad x \geq 0, \ t \geq 0,  
\ee
where the subscripts $x$ and $t$ denote partial differentiation and $\rho$ is the constant mass density per unit reference length. In addition, $\sigma$ and $\gamma$ are related by the constitutive relation
\be
\label{eq-ny-2x}
\sigma = W'(\gamma),  
\ee
where $W$ is the strain energy per unit reference length\footnote{Therefore $\sigma$ has the dimension of force.}.

Suppose that the motion involves a shock wave (whose image in the reference configuration is) at $x=s(t)$.  The displacement field is continuous at the shock but the stress, strain and particle speed are permitted to be discontinuous, with their limiting values satisfying the  jump conditions
\be
\label{eq-ny-6}
\sigma^+ - \sigma^- + \rho \sdot (v^+ - v^-) = 0, \qquad v^+ - v^- + \sdot (\gamma^+ - \gamma^-) = 0.  
\ee
Here $\sdot \coloneqq ds/dt$ is the shock speed and $g^+$ and $g^-$ denote the limiting values of a generic field $g(x, t)$ from $x=s(t)+$ and $x=s(t)-$ respectively. The limiting values must also obey the dissipation inequality\footnote{Equation \eqref{eq-20210512-3} shows, within the context of a particular problem, why ${\Bbb D}$ is the dissipation rate.}
\be
\label{eq-ny-7}
{\Bbb D} \coloneqq f \sdot \geq 0,  
\ee
where the driving force $f$ is 
\be
\label{eq-ny-8}
f \coloneqq W(\gamma^+) - W(\gamma^-) \ - \ \frac{\sigma^+ + \sigma^-}{2} (\gamma^+ - \gamma^-) ;  
\ee
e.g., see  \cite{RAJKK2006, Trusk1982}. It follows from  \eqref{eq-ny-6} that the shock speed can be expressed as
\be
\label{eq-ny-11x}
\sdot = \pm \sqrt{ \frac 1 \rho\, \frac{\sigma^+ - \sigma^-}{\gamma^+ - \gamma^-}} .  
\ee
Dissipation in an elastic material is only possible in the presence of a changing reference configuration due to the motion of, say, a singularity  such as  a shock wave, phase boundary or crack tip.  

The three problems to be studied in this paper will be described in the next and subsequent sections.  In order to compare the exact solution to Problem 1 with the numerical solution to Problem 2 and the approximate solution to Problem 3,  we now introduce the particular elastic material characterized by
\be
\label{eq-ny-2}
W(\gamma) = \frac 12 \mu \gamma^2 + \frac 16 \alpha^2 \gamma^3, \qquad \sigma = W'(\gamma) = \mu \gamma + \frac 12 \alpha^2 \gamma^2,\qquad \mu>0, \ \alpha \neq 0, 
\ee
where we shall only be concerned with positive strains $\gamma >0$.  The stress-strain curve corresponding to \eqref{eq-ny-2}$_2$ rises monotonically and is convex. We take it to be monotonic so as to avoid phase transition fronts and convex so that the shocks are admissible according to the Oleinik criterion \cite{Oleinik-chord-1959} as well as the dissipation inequality \eqref{eq-ny-7}.  For this material,  the driving force \eqref{eq-ny-8} takes the explicit form
\be
\label{eq-ny-10}
f  = \frac{\alpha^2}{12}\, (\gamma^- - \gamma^+)^3 , 
\ee
and  the shock speed \eqref{eq-ny-11x} can be written as
\be
\label{eq-ny-11}
\frac{\sdot}{c_0} = \pm \sqrt{ 1 +   \frac{\gamma^+ + \gamma^-}{2\beta_0^2}} , 
\ee
where we have set
\be
\label{eq-ny2-2}
\beta_0 \coloneqq \sqrt{{\mu}/{\alpha^2}}, \qquad c_0 \coloneqq \sqrt{\mu/ \rho} \ ; 
\ee
$c_0$ is the acoustic speed in the reference configuration. 
The dissipation inequality \eqref{eq-ny-7} with \eqref{eq-ny-10} implies that we should take the positive square root in \eqref{eq-ny-11} if $\gamma^- > \gamma^+$ and the negative square root in the opposite case.


\subsection{Problem 1.} \label{sec-1b}

Problem 1 concerns the aforementioned elastic bar.  The bar is unstressed and at rest at the initial instant $t=0$ and its free-boundary $x=0$ is subjected to a constant ``pulling'' speed $V$ for all time $t>0$. Thus we are concerned with the initial and boundary conditions
\be
\label{eq-20210516-1}
\gamma(x, 0) = 0, \quad v(x,0) = 0 , \quad x> 0 \qquad \mbox{and} \qquad
v(0,t) = - V, \quad t > 0. 
\ee
We shall refer to $V>0$ as the ``impact speed''.

For a material whose stress-strain relation increases monotonically and is convex\footnote{If the stress-strain relation is monotonic and concave, the strain and particle speed vary continuously and the solution involves a fan, $\gamma = \gamma(x/t), v = v(x/t)$, connecting two constant states.},
the solution to this problem has the piecewise constant form
\be
\label{eq-20210512-2}
\gamma(x, t), v(x, t) = \left\{
\ba{cll}
\gamma^-, \ -V, \qquad &0 < x < \sdot t,\\[2ex]

0,\  0, \qquad &x > \sdot t,\\
\ea
\right. 
\ee
involving a shock wave at $x=\sdot t$ that moves into the undisturbed material at a constant speed $\sdot$.   Thus the particle $x$  remains unstrained and at rest for times $0 < t < x/\sdot$; its strain and speed jump instantaneously to the values $\gamma^-$ and $-V$ as the shock passes through this point; and they remain at those values for $t > x/\sdot$. The shock speed, $\sdot$, and the strain behind the shock, $\gamma^-$, are to be determined.

The parameters $\sdot$ and $\gamma^-$  can be determined from the jump conditions \eqref{eq-ny-6} with $\gamma^+=0, v^+=0$. For the constitutive relation \eqref{eq-ny-2}, they tell us that the shock speed $\sdot(V)$ is the real positive root of the cubic equation
\be
\label{eq-20210515-1}
\left( \frac{\sdot}{c_0} \right)^3 - \frac{\sdot}{c_0}  - \frac 1{2 } \frac{V}{c_0\beta_0^2} = 0, 
\ee
and that the strain $\gamma^-(V)$ behind the shock is related to the impact speed $V$ through either of the equivalent expressions 
\be
\label{eq-20210510-1}
\gamma^-(V) = \frac{V}{\sdot(V)} , \qquad \qquad \frac{V}{c_0\beta_0^2} = \frac{\gamma^-(V)}{\beta_0^2} \, \sqrt{1 + \frac 12 \frac{\gamma^-(V)}{\beta_0^2}}. 
\ee
The relation \eqref{eq-20210510-1}$_2$ between $\gamma^-(V)$ and $V$ is monotonic and so there is a {one-to-one relation between the impact speed and the strain behind the shock}. A graph of $\gamma^-(V)$ versus $V$ will be displayed in Section \ref{sec-2}.

Let $X$ be an arbitrary fixed point in the bar and limit attention to times $t < X/\sdot$ so that this point lies {\it ahead} of the shock wave. Then it can be readily shown from \eqref{eq-20210512-1}, \eqref{eq-ny-2x}, \eqref{eq-ny-6} and \eqref{eq-20210516-1}$_3$ that
\be
\label{eq-20210512-3}
\sigma(0,t)\, V = \frac{d}{dt}\int_0^X E(x, t) \, dx \ +  \  \Bbb D,\qquad E(x, t) \coloneqq  \frac 12 \rho v^2(x, t) + W(\gamma(x, t)), 
\ee
where $\Bbb D$ is given by \eqref{eq-ny-7}, \eqref{eq-ny-8}.  The left-hand side of \eqref{eq-20210512-3}$_1$ represents the rate of external working on the segment $[0,X]$ of the bar and the first term on its right-hand side is the rate of increase of the kinetic plus potential energy of this segment. Therefore $\Bbb D$ represents the rate of dissipation. The fact that ${\Bbb D} \neq 0$ is due to the presence of the shock wave within the interval $[0,X]$. From \eqref{eq-ny-7}, \eqref{eq-ny-10}, \eqref{eq-ny-11} with  $\gamma^+=0$, the dissipation rate ${\Bbb D} = f \sdot$ in Problem 1 can be written as
\be
\label{eq-20210511-5}
\frac{\Bbb D}{\mu c_0 \beta_0^4}  = \frac{1}{12}   \,  \left(\frac{\gamma^-}{\beta_0^2}\right)^3  \sqrt{1 + \frac 12 \frac{\gamma^-}{\beta_0^2}  }. 
\ee
Plots of $\Bbb D$ versus $V$ will be presented in Sections \ref{sec-2} and \ref{sec-3}.


\section{Impact problem for a discrete system of particles.} \label{sec-2}

We now consider a semi-infinite chain of identical particles numbered  $j = 0, 1, 2, \ldots$, each of mass $m$. The $j$th particle is located at $x_j = h j$ in a reference configuration and at 
$y_j(t)$ at time $t$. Each particle interacts with its nearest neighbors (only) through identical nonlinear elastic springs. We shall refer to the spring connecting the $j$th and $j+1$th particles as the $j$th spring. The particle speed, $v_j$, and the elongation of the $j$th spring, $\delta_j$, are 
\be
\label{eq-20210517-1}
v_j = \dot y_j, \qquad \delta_j = y_{j+1} - y_{j} - h.  
\ee
If $U(\delta_j)$ denotes the potential energy of the $j$th spring,  the force in that spring is
\be
\label{eq-20210517-2}
\sigma_j = U'(\delta_j),  
\ee 
and a motion of the particle chain is described by the system of equations
\be
\label{eq-20210511-1}
m \dot v_j = U'(\delta_{j}) - U'(\delta_{j-1}), \qquad
\dot \delta_j = v_{j+1} - v_{j}. 
\ee

In order to compare the solutions of the discrete and continuous systems, we let
\be
\label{eq-20210517-3}
\gamma_j(t) \coloneqq \delta_j(t)/h,  
\ee 
be the strain in the $j$th spring and
introduce the energy per unit reference length, $W$, expressed as a function of strain:
\be
\label{eq-20210514-1}
W(\gamma_j) \coloneqq U(h\gamma_j)/h.  
\ee
It follows that the force in the $j$th spring is $
\sigma_j = U'(\delta_j) = W'(\gamma_j)$ 
where the prime denotes differentiation with respect to the argument. 
We also let
\be
\label{eq-20210515-2} 
\rho = m/h.  
\ee

Let $E_j$ denote the total energy of the $j$th spring-particle pair, i.e. the kinetic energy of the $j$th particle plus the potential energy of the $j$th spring:
\be
\label{eq-20210513-1}
E_j \coloneqq \frac 12 m v_j^2 +   U(\delta_{j}).  
\ee
The following balance equation can be {\it derived} from \eqref{eq-20210511-1}: 
\be
\label{eq-20210511-2}
U'(\delta_{j}) v_{j+1} - U'(\delta_{j-1}) v_{j} = \frac{dE_j}{dt}, \qquad j = 1, 2,  \ldots  \ . 
\ee
Considering the $j$th spring-particle pair as a system, equation \eqref{eq-20210511-2} states that the rate-of-working of the external forces on this system equals the rate of increase of its energy.


\subsection{Problem 2.} \label{sec-2b}

Problem 2  is the discrete counterpart of Problem 1 and concerns the aforementioned chain of particles.  At the initial instant the particles are at rest and the springs are unstretched.  For all time $t>0$ the zeroth particle is subjected to a constant ``pulling'' speed $V >0$ (and we again refer to it as the ``impact speed''). Thus we are concerned with the  initial and boundary  conditions 
\be
\label{eq-20210511-3}
v_j(0) = 0, \ j = 1, 2, \ldots ;  \qquad \delta_j(0) = 0, \ j = 0, 1, 2, \ldots ; \qquad
v_0(t) = -V, \ t >0.  
\ee
The initial boundary-value problem \eqref{eq-20210511-1}, \eqref{eq-20210511-3} was solved numerically for a chain with $N$ particles for the material characterized by
\be
\label{eq-20210516-2}
U(\delta) = h\, \left[ \frac 12 \mu \left(\frac{\delta}{h}\right)^2 + \frac 16 \alpha^2 \left(\frac{\delta}{h}\right)^3\right], \qquad \mu >0, \alpha \neq 0;  
\ee
cf. \eqref{eq-20210516-2} with \eqref{eq-ny-2}$_1$, \eqref{eq-20210514-1}; the associated acoustic speed is 
\be
c_0 \coloneqq \sqrt{\mu h/m} \ \stackrel{\eqref{eq-20210515-2}}= \ \sqrt{\mu/\rho}.  
\ee
We used the standard integrators in MATLAB as well as a leap-frog integrator to compute the solution  and stopped calculations before any waves reached the remote end of the chain.

\begin{figure}[h]
\begin{center}
\includegraphics[scale=0.85]{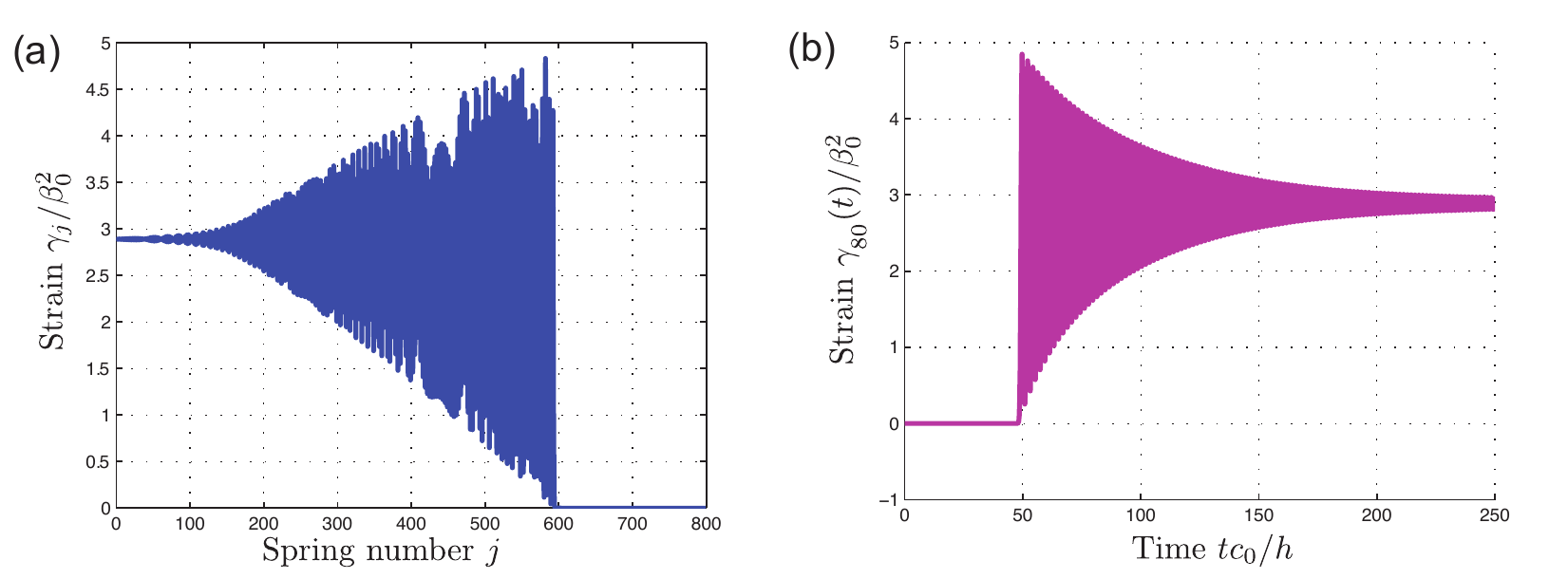}
\caption{\footnotesize $(a)$ Strain $\gamma_j(t)/\beta_0^2$ versus spring number $j$ at time $tc_0/h = 360$. $(b)$ Strain $\gamma_{80}(t)/\beta_0^2$  versus  time $tc_0/h$ at spring  $j=80$.  The value of the strain in a spring remains at zero for a certain initial period of time, undergoes a rapid increase  when the disturbance wave reaches it, and then undergoes rapid oscillations with slowly decaying amplitude. The amplitude of oscillation decreases linearly in $(a)$ and ``curvilinearly'' in $(b)$.  For these plots $V/(c_0\beta_0^2) = 4.48$, $N = 800$. }
\label{Fig-0.pdf}
\end{center}
\end{figure}

Figure \ref{Fig-0.pdf}  shows how the strain of the $j$th spring varies with the spring number $j$ (at a fixed time $t$) and with time $t$ (at a fixed spring $j$) in one (arbitrarily chosen) calculation.  The strain in a spring remains at the value zero for a certain initial period of time, undergoes a rapid increase  at some instant, and then undergoes rapid oscillations with slowly decaying amplitude.  Note that there are two time-scales involved: the slow time on which the amplitude decreases and the fast time on which the oscillations occur.  Observe also that the amplitude of oscillation as a function of $j$ decreases linearly (Figure \ref{Fig-0.pdf}$(a)$), whereas as a function of $t$ it decreases ``curvilinearly'' (Figure \ref{Fig-0.pdf}$(b)$). We shall revisit this observation in Section  \ref{sec-3}.

\begin{figure}[h]
\begin{center}
\includegraphics[scale=0.8]{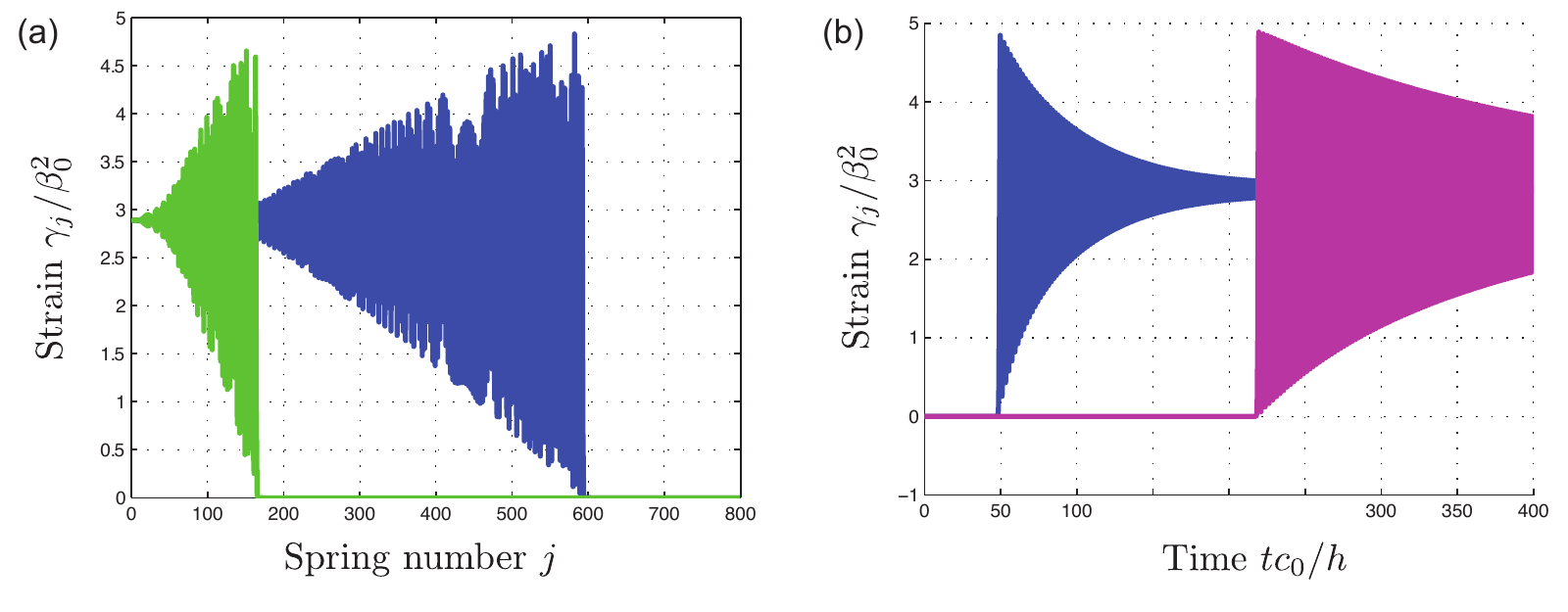}
\caption{\footnotesize Strain profiles: $(a)$ strain $\gamma_j(t)/\beta_0^2$ versus spring number $j$ at two times $tc_0/h = 100$ (left), $tc_0/h = 360$ (right).  $(b)$ Strain $\gamma_j(t)/\beta_0^2$ versus time $tc_0/h$ for two springs,  $j=80$ (left) and $j=360$ (right). For these plots, $V/(c_0\beta_0^2) = 4.48$, $N = 800$.}
\label{Fig-3.pdf}
\end{center}
\end{figure}

Figure \ref{Fig-3.pdf}  shows the results of a few such calculations. Observe that the solution involves a dispersive wave packet propagating into the quiescent material. The amplitude of oscillation at the leading edge remains constant as the wave packet propagates, but its width increases with time since the leading edge travels faster than the trailing edge. 

Several such calculations were carried out, and from them, we observed that for each spring $j$, 
\be
\big<\!\gamma_j(t) \!\big> \ \to \  \overline\gamma  \qquad \mbox{as} \quad t \to \infty;   
\ee
i.e. the strain $\gamma_j(t)$ in every spring $j$  approaches a value $\overline\gamma$ (independent of $j$) in the sense of a weak limit, meaning that the strain approaches an average value $\overline\gamma$ upon which are superposed  periodic oscillations.  In this paper, whenever we say that some quantity approaches a certain value, it will always be in this sense of a weak limit unless explicitly stated otherwise. The particle speed $v_j(t)$ similarly approaches the value $v_0=-V$ at each $j$ where $V$ is the impact speed.

\begin{figure}[h]
\begin{center}
\includegraphics[scale=0.75]{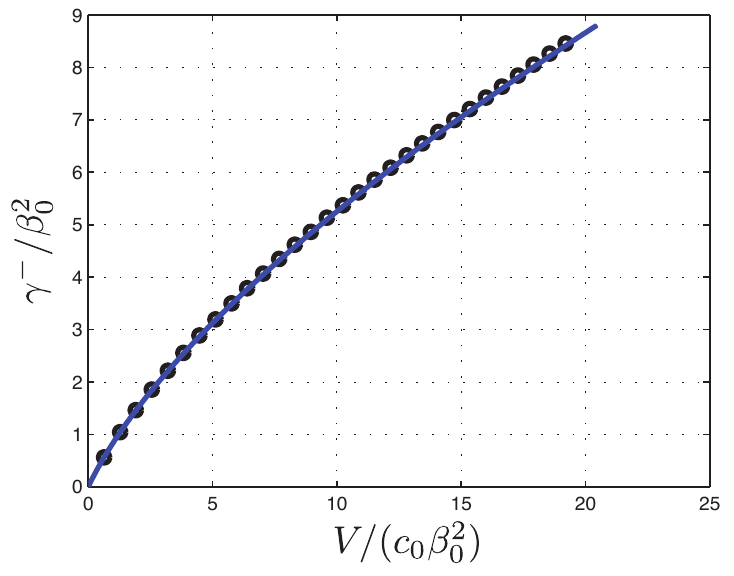}
\caption{\footnotesize The circles correspond to the limiting strain $\overline\gamma/\beta_0^2$ in Problem 2 as determined numerically while the solid curve corresponds to the strain $\gamma^-/\beta_0^2$ behind the shock in Problem 1 according to \eqref{eq-20210510-1}$_2$. 
}
\label{Fig-1.pdf}
\end{center}
\end{figure}

The limiting strain value $\overline\gamma$ is independent of spring number and time but depends on the impact speed.  Since  $\overline\gamma$ is found by solving a different set of equations to those in Problem 1, it is not a priori necessary that it equal the strain $\gamma^-$ behind the shock wave in Problem 1.  The circles in Figure \ref{Fig-1.pdf} show how $\overline \gamma$ varies with $V$ according to our numerical solution of Problem 2.  The variation of the strain $\gamma^-$ in Problem 1   corresponds to the solid curve.  It is difficult to distinguish between the two from the figure. This is consistent with the former problem being the discrete counterpart of the latter. From hereon we shall write $\gamma^-$ for $\overline\gamma$.

\begin{figure}[h]
\begin{center}
\includegraphics[scale=0.75]{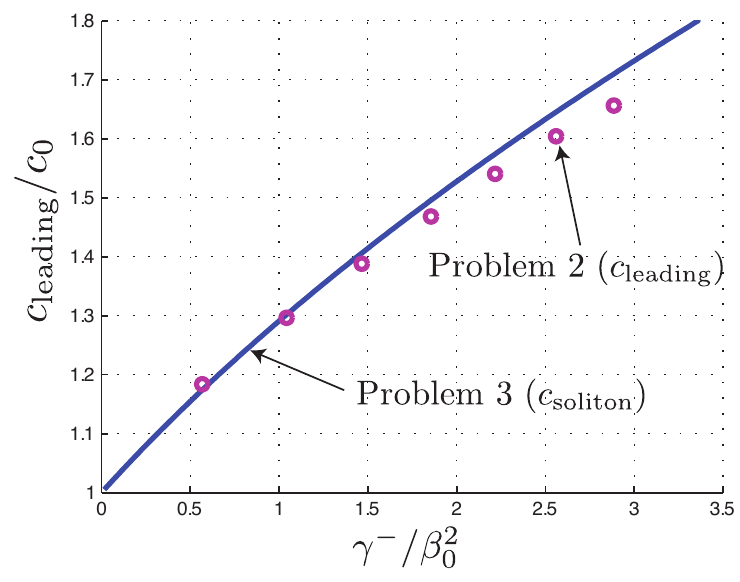}
\caption{\footnotesize The speed $c_{\rm leading}$ of the leading edge of the wave packet in Problem 2 (circles) determined numerically.  The solid curve corresponds to the speed $c_{\rm soliton}$ of the leading edge of the DSW in Problem 3 as will be given in  \eqref{eq-20210613-3}$_1$.}
\label{Fig-cleading.pdf}
\end{center}
\end{figure}

We next determine the speed of the leading edge of the propagating wave packet, $c_{\rm leading}$, or equivalently the spring number, $n(t)$, of the spring at the leading edge. This will be needed  in the next section. We identify the spring $n(t)$ at the leading edge using the criterion that it is the first spring in the chain whose strain has risen from $0$ and exceeded the (ad hoc) threshold value $1.0 \times 10^{-8}$. The speed of the leading edge is then given by $c_{\rm leading} = \dot n h$. We estimated $c_{\rm leading}$ for various values of the impact speed $V$ (or equivalently the strain $\gamma^-$). The results correspond to the circles in Figure \ref{Fig-cleading.pdf}. The solid curve there represents the speed, $c_{\rm soliton}$, of the leading edge of the wave packet in Problem 3 as will be derived later, see \eqref{eq-20210613-3}$_1$.

The  chain of particles connected by nonlinear elastic springs is a conservative system.  In fact, upon summing \eqref{eq-20210511-2}, one is led to
\be
\label{eq-3s}
\sigma_0 V = \frac{dE}{dt}  , \qquad  E(t) \coloneqq \sum_{j=0}^\infty E_j(t),  
\ee
where $\sigma_0(t) = U'(\delta_0(t))$ is the externally applied force on the zeroth particle\footnote{Since the zeroth particle travels at constant speed, the resultant force on it vanishes and therefore the externally applied force on it equals the force in the zeroth spring.}; $v_0(t)=-V$ is its speed; $\dot E_0 = U'(\delta_0)\dot \delta_0$ which follows from \eqref{eq-20210513-1} with $\dot v_0 = 0$; and
$E$ is the total energy in the system\footnote{We assume that the infinite sum in \eqref{eq-3s}$_2$ converges for the particular motions involved in Problem 2.}.   Equation \eqref{eq-3s} is simply a statement of the usual elastic power identity (``conservation of energy'') and should be compared with the corresponding equation \eqref{eq-20210512-3} for the elastic bar which involves an additional dissipative term.

While the theory implies that \eqref{eq-3s}
must necessarily hold, not all numerical schemes conserve energy.  In fact, the decaying strain amplitudes is Figures \ref{Fig-0.pdf} and \ref{Fig-3.pdf} are reminiscent of the oscillations of a {\it damped} system. For our purposes, where the calculation of energy underlies the central question being investigated, it is important that the discreteness of the particle chain not introduce any numerical dissipation. As described in section S1 of the electronic supplemental material, 
we confirmed that the numerical schemes used conserved energy and obeyed \eqref{eq-3s}.


\subsection{Oscillatory energy. Apparent dissipation.} \label{sec-2c}

Energy is not conserved in Problem 1 because of the propagating shock wave, while energy is conserved in its discrete counterpart Problem 2.  One way to heuristically understand the dissipation in Problem 1 in terms of the energy in Problem 2 is as follows: the strain and speed of all particles in the chain eventually settle at the values $\gamma^-$ and $v^-=-V$. This motivates us to introduce
\be
\label{eq-sz-1}
v^{\rm osc}_j(t) \coloneqq v_j(t) - v^-, \qquad \gamma^{\rm osc}_j(t) \coloneqq \gamma_j(t) - \gamma^-\qquad {\rm for} \quad 0 \leq j \leq n(t),
\ee
where  $n(t)$ is the particle at the leading edge of the propagating wave packet  at  time $t$.  We define  
the energy associated with the oscillatory part of the motion by
\be
\label{eq-sz-2}
E_{\rm osc}(t) \coloneqq \sum_{j=0}^{n(t)} \left(\frac 12 m \big(v^{\rm osc}_j(t)\big)^2 + hW(\gamma^{\rm osc}_j(t))\right),
\ee
and refer to it as the oscillatory or excess energy in the system.  Then the rate of increase of the oscillatory energy is
\be
\label{eq-sz-3}
D(t) = \frac{d}{dt}E_{\rm osc}(t).
\ee

We calculated $D$ using \eqref{eq-sz-3} as follows:  for each impact speed $V$,  we calculated the oscillatory energy $E_{\rm osc}(t)$ using the numerical solution to the problem together with  \eqref{eq-sz-1} and \eqref{eq-sz-2}; the particle $n(t)$ at the leading edge was determined using $\dot n(t) = c_{\rm leading}/h$ as described in Section \ref{sec-2b}. 
We then plotted $E_{\rm osc}(t)$ versus $t$ and observed that the relationship was linear (with small superposed jagged oscillations). We identified $D$ with the slope of this line which is effectively an averaging  over the rapid oscillations\footnote{A figure in the section S2 of the electronic supplemental material shows a graph of $E_{\rm osc}(t)$ versus $t$.}. Several such calculations were carried out for different values of the impact speed $V$.

\begin{figure}[h]
\begin{center}
\includegraphics[scale=0.75]{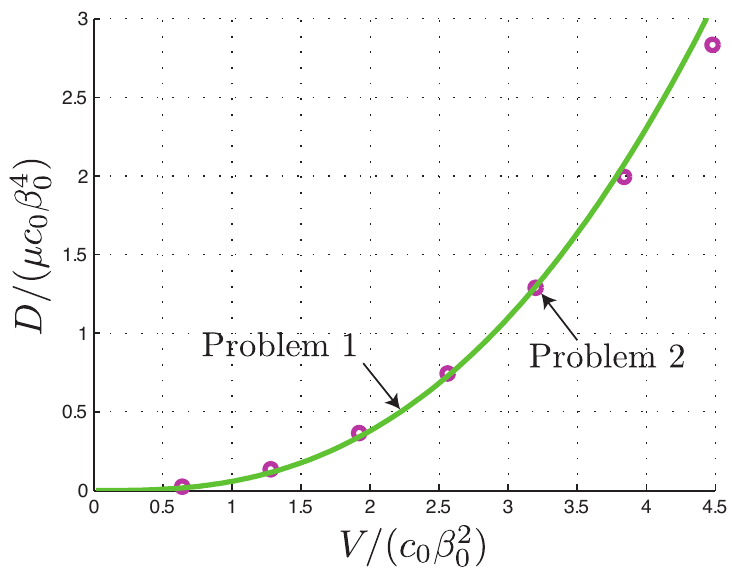}
\caption{\footnotesize The circles correspond to the oscillatory energy ${D}/(\mu c_0\beta_0^4)$ in Problem 2 while the solid curve corresponds to the dissipation-rate 
${\Bbb D}/(\mu c_0\beta_0^4)$ in Problem 1 according to \eqref{eq-20210511-5}, \eqref{eq-20210510-1}$_2$.}
\label{Fig-DissProb2.pdf}
\end{center}
\end{figure}

The circles in
 Figure \ref{Fig-DissProb2.pdf} show the variation of $D$  with the impact speed $V$ in Problem 2.  The dotted curve corresponds to the dissipation-rate $\Bbb D$ in Problem 1 as given by \eqref{eq-20210511-5}.  The results indicate that  the dissipation rate at the shock in Problem 1 is well modeled by  the rate of increase of the oscillatory energy in the conservative wave in Problem 2 (though they are not identical).

\section{Impact problem for a dissipationless dispersive continuum. } \label{sec-3}

Since the solution to the impact problem for the discrete chain (Problem 2) displays dispersion but no dissipation, we now turn to a continuum model that has these same two characteristics. Specifically, we add a strain-gradient term to the stress-strain relation \eqref{eq-ny-2}$_2$ of Problem 1, and thus take the  constitutive relation of the one-dimensional semi-infinite continuum to be 
\be
\label{eq-20210612-3}
\sigma = W'(\gamma) + \eta h^2 \gamma_{xx}, \qquad W(\gamma) = \frac 12 \mu \gamma^2 + \frac 16 \alpha^2 \gamma^3;  
\ee
here $\eta >0$ and $h >0$ are constant parameters.

Taking $\eta >0$ in \eqref{eq-20210612-3}  is motivated by Taylor expanding the discrete equations of Problem 2 for small $h$; e.g. $\eta/\mu = 1/12$ according to Rosenau \cite{Rosenau1986}.  However, $\eta >0$ leads to instability at perturbations whose wave lengths are smaller than some critical value.  As noted by Sharma \cite{Sharma2005},  despite this deficiency, interesting results can be derived in this case, \cite{Kresse2003, RASV1999, RASV2003}. In the context of the present problem,  we show in the appendix that if we limit attention to perturbations whose wave lengths remains close to the wave lengths of the solutions encountered here, then stability is maintained.

Consider a motion $y(x, t) = x + u(x, t)$
where $u(x, t)$ is the displacement of particle $x$ at time $t$. We do {\it not} assume $u$ or its derivatives to be small. The strain and particle speed associated with this motion are $\gamma = y_x - 1 = u_x, v = y_t = u_t$.
Substituting  \eqref{eq-20210612-3}$_1$ into the equation of motion $\sigma_x = \rho v_t$, and from the definitions of strain and particle speed, one obtains the following pair of partial differential equations for $\gamma(x.t), v(x, t)$:
\be
\label{eq-20210612-5}
\eta_0 h^2 \gamma_{xxx} + \gamma_{x} + \beta_0^{-2} \gamma \gamma_{x} = c_0^{-2} v_{t} , \qquad  v_x = \gamma_t. 
\ee
Here we have set
$$
\eta_0 = \eta/\mu, 
$$
and $c_0 = \sqrt{\mu/\rho}$ and $\beta_0 = \sqrt{\mu/\alpha^2}$ as before.

Now consider traveling wave solutions of the form
\be
\label{eq-20210524-3}
\gamma(x, t) = g(\Phi), \quad v(x, t) = w(\Phi), \qquad \Phi = \frac{kx - \omega t}{h}, \qquad c = \omega/k, 
\ee
where the wave number $k$, frequency $\omega$ and phase speed $c$ are constants (for the moment). From  \eqref{eq-20210612-5} and \eqref{eq-20210524-3} we obtain 
\be
\label{eq-20210617-1}
\eta_0 k^2 g''' + g' + \beta_0^{-2} gg' = -(c/c_0^2) w', \qquad w' = -c g'. 
\ee
Integrating \eqref{eq-20210617-1}$_2$ gives $w(\Phi) = - c g(\Phi) + v_*,$
where $v_*$ is a constant (for the moment).   Thus, and by substituting \eqref{eq-20210617-1}$_2$ into \eqref{eq-20210617-1}$_1$, we can rewrite \eqref{eq-20210617-1} as the following pair of equations for $g(\Phi)$ and $w(\Phi)$:
\be
 \label{eq-0406-3}
\eta_0 k^2 g''' - (c^2/c_0^2 -1)  g' + \beta_0^{-2} gg' = 0, \qquad w(\Phi) =  v_*- c\, g(\Phi) . 
\ee
 The strain and particle speed in the traveling wave can now be expressed as
\be
 \label{eq-20210612-6}
\gamma(x, t) =  g(\Phi), \qquad v(x, t) = v_* - c \, g(\Phi), \qquad \Phi = \frac{kx - \omega t}{h}. 
\ee
Once a traveling wave solution for the strain is determined from  \eqref{eq-0406-3}$_1$, the associated traveling wave for the particle speed is given immediately by \eqref{eq-0406-3}$_2$ to within the arbitrary constant $v_*$.


\subsection{Steady periodic traveling wave.}

 Based on Figures \ref{Fig-0.pdf} and \ref{Fig-3.pdf}, the solution to the impact problem for the discrete system involves a modulated traveling wave packet in which the amplitude of oscillation decays slowly, much more slowly than the time-scale associated with the frequency of oscillation. In order to construct such a solution we follow the approach introduced by Whitham \cite{Whitham1965a, Whitham1970} that has since been further developed and used by many authors, e.g. see the review article by El and Hoefer \cite{elhoefer}, the dissertation by Nguyen \cite{Nguyen}, the book by Kamchatnov \cite{Kamchatnov2000} and the references therein.  
The procedure is to first construct an exact {\it periodic} traveling wave solution, and to then allow the parameters in that solution to vary slowly in an appropriate manner. For example, the periodic traveling wave may have the form $\widehat\gamma((kx-\omega t)/\veps, p)$ where $k, \omega$ and $p$ are constant parameters, with the modulated wave having  the form  $\widehat\gamma(\theta(x, t)/\veps, p(x, t))$ where $\theta(x, t)$ and $p(x, t)$ are slowly varying functions; here $\veps = h/L \ll 1$ where  $L$ is a macroscopic length involved in the problem.  Such waves involve two slow scales $x$ and $t$ and two fast scales $x/\veps$ and $t/\veps$. 
When $\veps=0$, the underlying system of partial differential equations is hyperbolic and its solution can involve a shock wave (as in Problem 1). The term $\veps > 0$ introduces dispersion into the problem (but not dissipation) and the solution corresponding to a shock wave is referred to as a  {\it dispersive shock wave} (DSW), e.g. \cite{hoeferscholarpedia}.

Integrating  \eqref{eq-0406-3}$_1$ twice leads to
 \be
 \label{eq-20210523-1}
 \big(g'\big)^2 = \frac{1}{3 \kappa^2} \Big[d_1 + d_2 g + 3 \beta_0^2 (c^2/c_0^2 - 1) g^2 - g^3\Big],
 \ee
 where $d_1$ and $d_2$ are constants of integration and
 \be
 \label{eq-20210519-12} 
 \kappa \coloneqq  k \beta_0 \sqrt{\eta_0}.  
 \ee
With the exception of the coefficient in front of the term $g^2$,  equation \eqref{eq-20210523-1} is the same equation that is arrived at when analyzing the Korteweg-de Vries (KdV) equation. We shall therefore simply write down the relevant solution of \eqref{eq-20210523-1} and list its key features without derivation and refer the reader to the literature on DSWs in the KdV equation
 for details, e.g. Section IV-B of \cite{hoefer-2006}.

A three-parameter family of $2\pi$-periodic solutions of \eqref{eq-20210523-1} is
\be
\label{eq-20210519-11}
g(\Phi) = g^- - m^2 (g^- - g^+) + 2 m^2(g^- - g^+) {\rm cn}^2\left(\frac{K(m)}{\pi} \, \Phi; m\right),    
\ee
where ${\rm cn}(z,m)$ is a Jacobi elliptic function\footnote{Definitions and properties of this and the other elliptic functions encountered in this paper can be found, for example, in  \cite{ref-byrd-elliptic, ref-NIST:DLMF}. It should be noted that the parameter we call $m^2$ is taken by some authors, including MATHEMATICA, to be $m$. },  $K(m)$ is the complete elliptic integral of the first kind, and the three constant parameters  
$g^-, g^+$ and $m$ are arbitrary except for the requirements
$$
g^- > g^+, \qquad 0 \leq m \leq 1. 
$$
The associated phase speed $c$, wave number $k$ and group speed $V_g$ are
\be
\label{eq-0406-11}
\frac{c}{c_0} = \sqrt{ 1 +   \left[ \frac{2g^+ + g^- + m^2 \,(g^- - g^+)}{3\beta_0^2}\right] }\, ,  
\ee
\be
\label{eq-0406-10}
k =  \frac{\pi}{K(m)}\,  \sqrt{\frac{g^- - g^+}{6\eta_0\beta_0^2}} ,   
\ee
\be
\label{eq-0406-44}
\frac{V_g}{c_0} = \frac{c_0}{c} \left[ \frac{c^2}{c_0^2} -  \frac{mK(m)}{K'(m)}\, \frac{g^- - g^+}{3\beta_0^2}\right]  , 
\ee
where $K'(m)$ is the derivative of $K(m)$ with respect to $m$.  The function ${\rm  cn}^2[\cdot,m]$ oscillates between the values $0$ and $1$ and so the (peak to valley) amplitude of oscillation in \eqref{eq-20210519-11} is
\be
\label{eq-20210519-13}
a = 2 m^2(g^- - g^+).  
\ee
The three parameters $g^+, g^-$ and $m$ can of course be replaced by the three ``physical parameters'', phase speed $c$, wave number $k$ and amplitude $a$. Note that the amplitude, phase speed and group speed do not depend on the strain-gradient parameter $\eta$ but the wave number does.

It will be useful for future purposes to note that the average of $g(\Phi)$ over the oscillations, defined by 
\be
\label{eq-20210701-1}
\big< g \big> \coloneqq \frac{1}{2\pi} \int_0^{2\pi} g(\Phi)\, d\Phi ,
\ee
is
\be
\label{eq-20210520-1}
\big< g \big> = 2 g^+ - g^- + m^2(g^- - g^+) + 2(g^- - g^+) \frac{E(m)}{K(m)} ,  \quad 
\ee
where $E(m)$ is the complete elliptic integral of the second kind.

Turning next to the particle speed, the periodic traveling wave solution is obtained immediately by substituting \eqref{eq-20210519-11}  into   \eqref{eq-0406-3}$_2$ which gives
\be
\label{20210620-eq3}
w =  w^- - m^2 (w^- - w^+) + 2 m^2(w^- - w^+) {\rm cn}^2\left(\frac{K(m)}{\pi} \, \Phi; m\right),  
\ee
where
\be
\label{20210620-eq3b}
w^- = v_* - cg^-, \qquad w^+ = v_* - cg^+.
\ee
This involves four constant parameters, three of which ($g^-, g^+$ and $m$) are the same as in the solution for the strain. The fourth parameter $v_*$ is an additional arbitrary constant. The phase speed $c$ appearing in \eqref{20210620-eq3b} is known in terms of $g^-, g^+$ and $m$; see \eqref{eq-0406-11}.  The average value of $w(\Phi)$ is 
\be
\label{eq-20210627-1}
\big<w\big>   =v_* - c \, \big< g \big>  = 2 w^+ - w^- + m^2(w^- - w^+) + 2(w^- - w^+) \frac{E(m)}{K(m)}.
\ee



\subsection{Slow modulation of the periodic traveling wave solution. Dispersive shock wave (DSW).} \label{sec-3b}

For our purposes in this paper, it is not necessary  that we construct the most general slow modulation of  \eqref{eq-20210519-11}.  Therefore, while we could specialize what is known for the KdV equation, it is easier to  tackle our problem directly instead. Moreover, as already mentioned in the Introduction, our analysis of the system of equations governing the motion of a one-dimensional continuum in its Lagrangian form is in fact simpler than the analysis of its Eulerian counterpart as can be found in the literature on DSWs in one-dimensional compressible fluid flows, e.g. Section IV-D of \cite{hoefer-2006}.  For both these reasons we shall provide some details of the calculations to follow.  


\subsubsection{Strain field $\gamma(x, t)$.} \label{sec-3b1}

We first construct a slow modulation of the preceding periodic traveling wave solution for the strain field by allowing (one or more of) the three parameters $g^-, g^+$ and $m$ in   \eqref{eq-20210519-11} to be slowly varying functions of $x$ and $t$.

First consider the parameter $m$ that is required to be in the range $0\leq m \leq 1$.  If $m(x, t)$ varies from $0$ to $1$ as one moves from the trailing edge to the leading edge of the wave packet, according to \eqref{eq-20210519-13}  the amplitude of oscillation would increase from $0$ to $2(g^- - g^+)>0$ (qualitatively as in Figure \ref{Fig-3.pdf}). Next, from \eqref{eq-20210520-1} and the properties of the complete elliptic integrals $E(m)$ and $K(m)$, 
\be
\label{eq-20210524-4}
 \big< g\big> \  \to\  g^- \quad \mbox{as} \quad m \to 0, \qquad   \big< g\big> \  \to\  g^+ \quad \mbox{as} \quad m \to 1,  
\ee
and so the average value of $g$ varies from $g^-$ to $g^+$ when $m$ varies from $0$ to $1$. In view of these observations, and since we will eventually be interested in a solution that connects two {\it constant states}, it is natural (though not necessary) to limit attention to the special case where the two parameters $g^-, g^+$ remain constant and only allow $m=m(x, t)$ to be slowly varying.  Observe from the relevant formulae in the preceding sub-section that the amplitude, wave number, group speed etc. are all functions of $m$ (but not $\Phi$) and so they will vary slowly.  We need to determine $m(x, t)$.

We make one more set of observations before turning to finding $m(x, t)$. Since $m$ ranges over the interval $[0,1]$, it is useful to look at the solution \eqref{eq-20210519-11} in the two limiting cases $m\to 0$ and $m \to 1$.   When $m \to 1$ one can show from \eqref{eq-0406-10} that $ k \to 0$ (so that the wave length $\to \infty$) and that $g(\Phi)$ is described by the soliton
$$
g(\Phi) =  g^+ + 2 (g^- - g^+) {\rm sech}^2\left(\frac{K(m)}{\pi}\, \Phi\right). 
$$
According to \eqref{eq-0406-11} with $m=1$, the soliton propagates at the particular phase speed
\be
\label{eq-0406-43}
\frac{c_{\rm soliton}}{c_0} \coloneqq \sqrt{ 1 +    \frac{g^+ + 2 g^-}{3\beta_0^2} } .   
\ee
In the  limit $m \to 0$ one sees  that $g(\Phi)$ is described by the constant solution
\be
\label{eq-20210521-12}
g(\Phi) = g^- .   
\ee
For small $m$ one has
$$
g(\Phi) \sim  g^-  + \frac a2 \cos \Phi, \qquad a = 2 m^2 (g^- - g^+),
    $$
 which is a harmonic wave propagating, according to \eqref{eq-0406-11}, at the phase speed
\be
\label{eq-0406-42}
\frac{c_{\rm harmonic}}{c_0} \coloneqq \sqrt{ 1 +    \frac{2g^+ + g^- }{3\beta_0^2} }. 
\ee

A curious factoid is that if one sets $g^+ = \gamma^+$ and $g^- = \gamma^-$ in \eqref{eq-0406-43} and \eqref{eq-0406-42}, one finds that the speed $\sdot$ of the shock wave in the elastic continuum (as given in \eqref{eq-ny-11}) is related to the phase speeds $c_{\rm soliton}$ and $c_{\rm harmonic}$ by
$$
\sdot^2  = \frac 12 ({c^2_{\rm soliton}}  + {c^2_{\rm harmonic}} ). 
$$

We now turn to determining the function $m(x, t)$, and for this we need another equation.  This equation, or more generally the three equations that would be needed had we permitted $g^-$ and $g^+$ to also vary, are obtained by either the singular perturbation method of two-timing, variational methods, or averaging three supplementary conservation laws,  \cite{Whitham1965a, Whitham1965b, Whitham1967, Whitham1970}.  One of the equations that typically arises from all such derivations is the so called {\it conservation of waves} equation, 
\be
\label{eq-20210520-2}
\frac{\partial \omega}{\partial x} + \frac{\partial k}{\partial t} = 0,    
\ee
relating the frequency and wave number, $\omega(x, t)$ and $k(x, t)$, of the modulated wave. We take for granted that \eqref{eq-20210520-2} is the requisite additional equation. Since $V_g = d\omega/dk$, this can  alternatively be written as
${\partial k}/{\partial t} + V_g \, {\partial k}/{\partial x} = 0 $
which is the usual statement that wave numbers propagate at the group speed.
Since $k(x, t)$ varies only due to the variation of $m(x, t)$, i.e. $k$ is a function of $m$, this in turn leads to
\be
\label{eq-0407-1}
\frac{\partial m}{\partial t} + V_{g}(m)\frac{\partial m}{\partial x} = 0, 
\ee
where $V_g(m)$ is the group speed given by \eqref{eq-0406-44}. Once \eqref{eq-0407-1} (with initial/boundary conditions as needed) has been solved for $m(x, t)$,
the solution $g(x, t)$ is given by \eqref{eq-20210519-11}.

Finally, in light of the particular problem we want to study, we restrict attention to the case where $m$ is scale-invariant so that $m(x, t) = m(x/t)$. Then \eqref{eq-0407-1} reduces to the algebraic equation
\be
\label{eq-0407-2}
\frac{x}{t} = V_{g}(m). 
\ee
Upon using \eqref{eq-0406-44} this can be written explicitly as
\be
\label{eq-0407-3}
\frac xt  = \frac{c_0^2}{c(m)} \left[ \frac{c^2(m)}{c_0^2} - \frac{1}{\beta_0^2} \frac{mK(m)}{K'(m)}\, \frac{g^- - g^+}{3}\right] , 
\ee
where the phase speed $c(m)$ is given by \eqref{eq-0406-11}. Equation \eqref{eq-0407-3} gives $x/t$ as a function of $m$, whose inverse yields $m=m(x/t)$.

Thus in summary, the strain field $\gamma(x, t)$ in the DSW is given by 
\be
\label{eq-20210622-1}
\gamma(x, t) = g^- - m^2 (g^- - g^+) + 2 m^2(g^- - g^+) {\rm cn}^2\left(\frac{K(m)}{\pi}\, \Phi; m\right),   
\ee
where $m=m(x/t)$ is determined by inverting \eqref{eq-0407-3}.  The parameters $g^+$ and $g^-$ are constants and $\Phi = k(m) (x - c(m) t)/h$ with $c(m)$ and $k(m)$ given by \eqref{eq-0406-11} and \eqref{eq-0406-10}. Figure \ref{Fig-13x.pdf} in Section \ref{sec-3c} shows the strain profile in Problem 3 based on \eqref{eq-20210622-1}, \eqref{eq-0407-3}.


\subsubsection{Particle speed $v(x, t)$:} \label{sec-3b2}

In order to construct the slowly modulated wave for the particle speed we turn to \eqref{20210620-eq3} with $w^\pm(x,t) = v_*(x,t) - c(m(x,t)) g^\pm$.  Since the modification to $m(x,t)$  has already been dealt with in the preceding sub-section, it remains to determine the slowly varying function $v_*(x, t)$.  We again restrict attention to the special case where $v_*(x, t)$  is scale invariant: $v_* = v_*(x/t)$. However, since $x/t = V_g(m)$, we may equivalently say that $v_* = v_*(m)$ whence we can write the particle speed field as
\be
\label{20210620-eq4}
v(x,t) =  w^-(m) - m^2 (w^-(m) - w^+(m)) + 2 m^2(w^-(m) - w^+(m)) {\rm cn}^2\left(\frac{K(m)}{\pi} \, \Phi; m\right),  
\ee
where
\be
\label{20210620-eq4b}
w^\pm(m) = v_*(m) - c(m)g^\pm, 
\ee
with $m(x/t)$ given by \eqref{eq-0407-3} and $v_*(m)$ to be determined. Observe that in the particular solution we have constructed, in contrast to $g^\pm$, the quantities $w^\pm(m)$ are not constants.

In order to find $v_*(x, t) = v_*(m)$ we average the conservation law\footnote{Section S5 of the electronic supplemental material gives the set of four conservations we could use had we allowed all four parameters $\gamma^+, \gamma^-, m$ and $v_*$ to be slowly varying.}  $v_x = \gamma_t$ over the fast oscillations (i.e. with respect to $\Phi$) to get
$$
\frac{\partial }{\partial x}\big<w\big>   = \frac{\partial }{\partial t} \big<g\big> ; 
$$
see \eqref{eq-20210701-1} for the definition of the average and note that, since the period, $2\pi$, of oscillation is constant, the averaging integral can be moved inside the derivatives. 
Since $\big<w\big>$ and $\big<g\big>$ depend on $x, t$ only through $m(x/t)$, this yields 
$\big<w\big>{}'  = - V_g \, \big<g\big>{}'$
where a prime denotes differentiation with respect to $m$ and we have used \eqref{eq-0407-1}.  On using $\big<w\big> = v_* - c \big<g\big>$, \eqref{eq-0406-11} and \eqref{eq-0406-44}  this leads to
$$
v_*'(m) = \frac{g^- - g^+}{3 \beta_0^2} \, \frac{c_0^2}{c(m)} \frac{m}{K'(m)} \frac{d}{dm}\Big( \big<g\big> K(m)\Big), 
$$
which can be further simplified using \eqref{eq-20210520-1} to
$$
v_*'(m) = \frac{g^- - g^+}{3 \beta_0^2} \, \frac{c_0^2}{c(m)} m\big[g^- - (g^- - g^+) m^2\big]. 
$$
Finally, this can be integrated (by changing the variable of integration from $m$ to $c$) to obtain
\be
\label{eq-20210621-10}
v_*(m) = v_*(1) +  (2 g^- + 2 g^+ + 3 \beta_0^2)(c-c_{\rm soliton}) + (\beta_0^2/c_0^2) (c_{\rm soliton}^3 - c^3), 
\ee
having used the fact that $c= c_{\rm soliton}$ when $m=1$.

Thus in summary, the particle speed field $v(x, t)$ in the DSW is given by \eqref{20210620-eq4},  \eqref{20210620-eq4b}, \eqref{eq-20210621-10}
where $g^+$, $g^-$ and $v_*(1)$ are constants;  $\Phi = k(m) (x - c(m) t)/h$ with $c(m)$ and $k(m)$ given by \eqref{eq-0406-11} and \eqref{eq-0406-10}; and $m=m(x/t)$ is determined by inverting \eqref{eq-0407-3}.


\subsection{Problem 3}\label{sec-3c}

We now use the preceding modulated traveling wave to construct an {\it approximate} solution to the impact problem for the dispersive continuum under consideration.  We refer to this problem as Problem 3.

\begin{figure}[h]
\begin{center}
\includegraphics[scale=0.5]{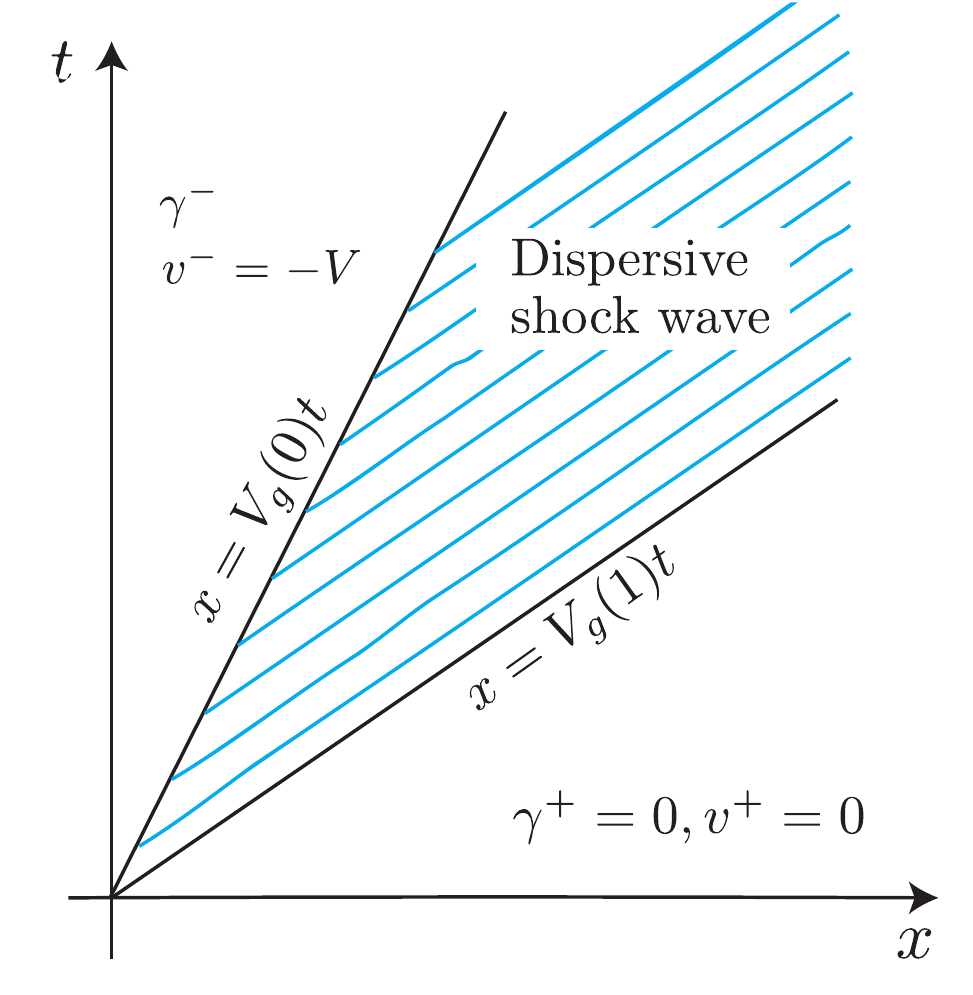}
\caption{\footnotesize The schematic $x, t$-plane associated with the approximate solution (Problem 3).}
\label{Fig-xtPlane.pdf}
\end{center}
\end{figure}

Recall that the strain and particle speed can be identified with
\be
\label{eq-20210526-1}
\gamma(x, t) = g(x, t), \qquad v(x, t) = w(x, t) = v_*(m) - c(m) g(x, t), \qquad m = m(x/t). 
\ee
We take for granted that the $x, t$-plane is as shown schematically in Figure \ref{Fig-xtPlane.pdf} where the strain and particle speeds ahead of and behind the DSW are constant and have the values, say, $\gamma^+, v^+$ and $\gamma^-, v^-$ respectively. (It is instructive not to take $\gamma^+ = 0, v^+=0$ initially though we shall do so later.) 

First consider the leading edge of the DSW.  Since $m=1$ here, it follows from \eqref{eq-0407-2}
that the leading edge is described by $x = V_g(1) t$,
and from \eqref{eq-20210524-4} and  \eqref{eq-0406-43} that $\big<\gamma\big> = g^+$ and $	c=c_{\rm soliton}$ there. Thus by this and \eqref{eq-20210526-1}, the average strain and particle speed just behind the leading edge are $g^+$ and $\big< v \big> = w^+(1)=v_*(1) - c_{\rm soliton}g^+$. Matching them to the strain and particle speed ahead of the leading edge thus gives
\be
\label{eq-20210525-12}
g^+ = \gamma^+, \qquad w^+(1) = v_*(1) - c_{\rm soliton}\gamma^+ = v^+.  
\ee

Similarly, since $m=0$ at the trailing edge, one has $x = V_g(0) t$,
$\big<\gamma \big> = g^-$ and $	c=c_{\rm harmonic}$ there. It therefore follows that the average strain and particle speed just inside of the trailing edge are $g^-$ and $\big< v \big> =w^-(0) = v_*(0) - c_{\rm harmonic}g^-$, and so,  matching across the trailing edge leads to
\be
\label{eq-20210525-13}
g^- = \gamma^- , \qquad w^-(0) = v_*(0) - c_{\rm harmonic} \gamma^- = v^- . 
\ee

From \eqref{eq-20210525-12}, \eqref{eq-20210525-13}, \eqref{eq-20210621-10}, \eqref{eq-0406-43} and \eqref{eq-0406-42} one obtains the following relation between $\gamma^\pm$ and $v^\pm$:
\be
\label{eq-20210623-1}
v^+ - v^-  =  2 (\beta_0^2/c_0^2) \left[  c_{\rm soliton}^3 - c_{\rm harmonic}^3 \right] ;
\ee
here $c_{\rm soliton}$ and $c_{\rm harmonic}$ are given by \eqref{eq-0406-43} and \eqref{eq-0406-42} respectively.
Equation \eqref{eq-20210623-1} is an explicit relation between the states $\gamma^-, v^-$ behind the DSW and the state $\gamma^+, v^+$ ahead of it. It is the counterpart of a Rankine-Hugoniot jump condition at a shock and the corresponding integral relation at a fan.

\begin{figure}[h]
\begin{center}
\includegraphics[scale=0.33]{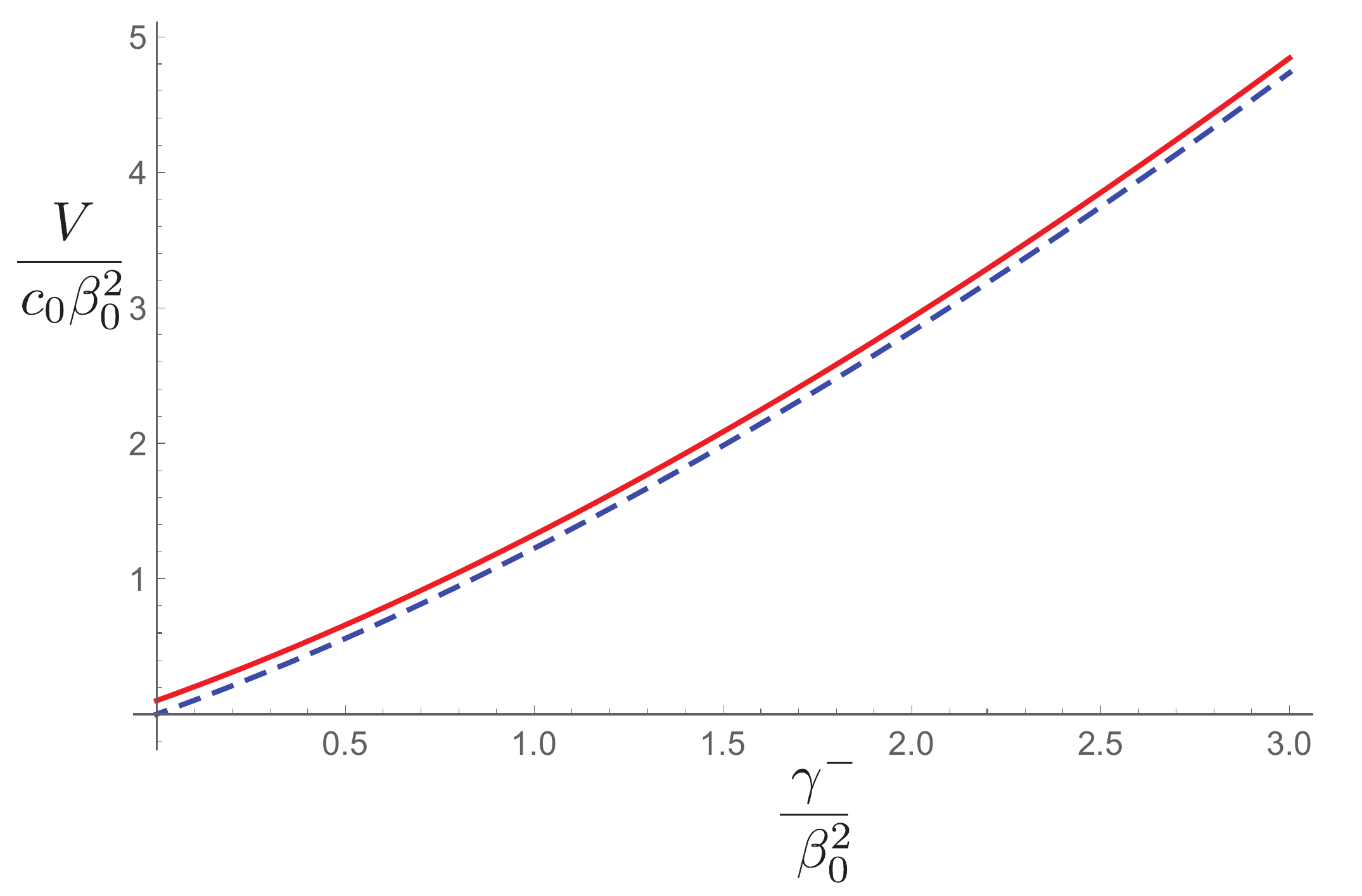}
\caption{\footnotesize The impact speed $V/(c_0\beta_0^2)$ versus the strain $\gamma^-/\beta_0^2$  for Problem 3 (dashed, eqn. \eqref{eq-20210621-12}) and for Problem 1 (solid, eqn.\eqref{eq-20210510-1}$_2$). The two curves fall on top of each other and for clarity we have shifted the curve corresponding to the shock (red) by 0.1 units vertically. The figure has been drawn for impact speeds conforming to \eqref{eq-20210621-13} below.}
\label{Fig-V-gammaminus}
\end{center}
\end{figure}

In the specific problem at hand, the system is quiescent initially and so $\gamma^+ = v^+ = 0$. Behind the wave packet we have $v^-=-V$ where $V$ is the impact speed.  On using this, \eqref{eq-20210623-1} simplifies to \
\be
\label{eq-20210621-12}
\frac{V}{c_0} =  2 \beta_0^2 \left[  \left(\frac{c_{\rm soliton}}{c_0}\right)^3 - \left(\frac{c_{\rm harmonic}}{c_0}\right)^3 \right] , 
\ee
where $c_{\rm soliton}$ and $c_{\rm harmonic}$ specialize to
\be
\label{eq-20210613-3}
\frac{c_{\rm soliton}}{c_0} \coloneqq  \sqrt{ 1 + \,  \frac{2 \gamma^- }{3\beta_0^2} } ,   
\qquad 
\frac{c_{\rm harmonic}}{c_0} \coloneqq  \sqrt{ 1 + \,   \frac{\gamma^- }{3\beta_0^2} }.  
\ee
This is an implicit algebraic equation for determining the strain $\gamma^-$ behind the DSW corresponding to the given impact speed. Figure \ref{Fig-V-gammaminus} shows a plot of $V$ versus $\gamma^-$ according to \eqref{eq-20210621-12}. For comparison we have also plotted the $V-\gamma^-$ relation \eqref{eq-20210510-1}$_2$ for the shock wave in Problem 1. The two curves fall on top of each other and for clarity we have shifted the curve corresponding to the shock (red) by 0.1 units vertically. It follows that the strain and particle speed behind the DSW in Problem 3 is essentially identical to the strain and particle speed behind the shock in Problem 1.  Recall from the discussion surrounding Figure \ref{Fig-1.pdf} that we previously made a similar observation between Problems 2 and 1.

In view of  $\gamma^+=0, v^+ = 0$ and \eqref{eq-20210525-12}$_2$, equation
 \eqref{eq-20210621-10} reduces to
\be
\label{eq-20210622-2}
v_*(x, t) = v_*(m) = - (\beta_0^2/c_0^2) \, (c_{\rm soliton} - c)^2 ( 2 c_{\rm soliton} + c)  .
\ee
where 
\be
\frac{c(m)}{c_0} = \sqrt{ 1 +    \frac{(1 + m^2) \,\gamma^- }{3\beta_0^2}  }\, .   
\ee

Thus in summary, given the impact speed $V$ and the constitutive parameters $c_0, \beta_0$,  we find $\gamma^-$ from  \eqref{eq-20210621-12}, 
 $m(x/t)$ from \eqref{eq-0407-3} and $v_*(x/t)$ from \eqref{eq-20210622-2}. The strain and particle speed fields within the DSW are then given by  \eqref{eq-20210622-1}, \eqref{20210620-eq4} and \eqref{20210620-eq4b}  with $g^+=0, g^- = \gamma^-$. The fields are constant on either side of the DSW.

\begin{figure}[h]
\begin{center}
\includegraphics[scale=0.3]{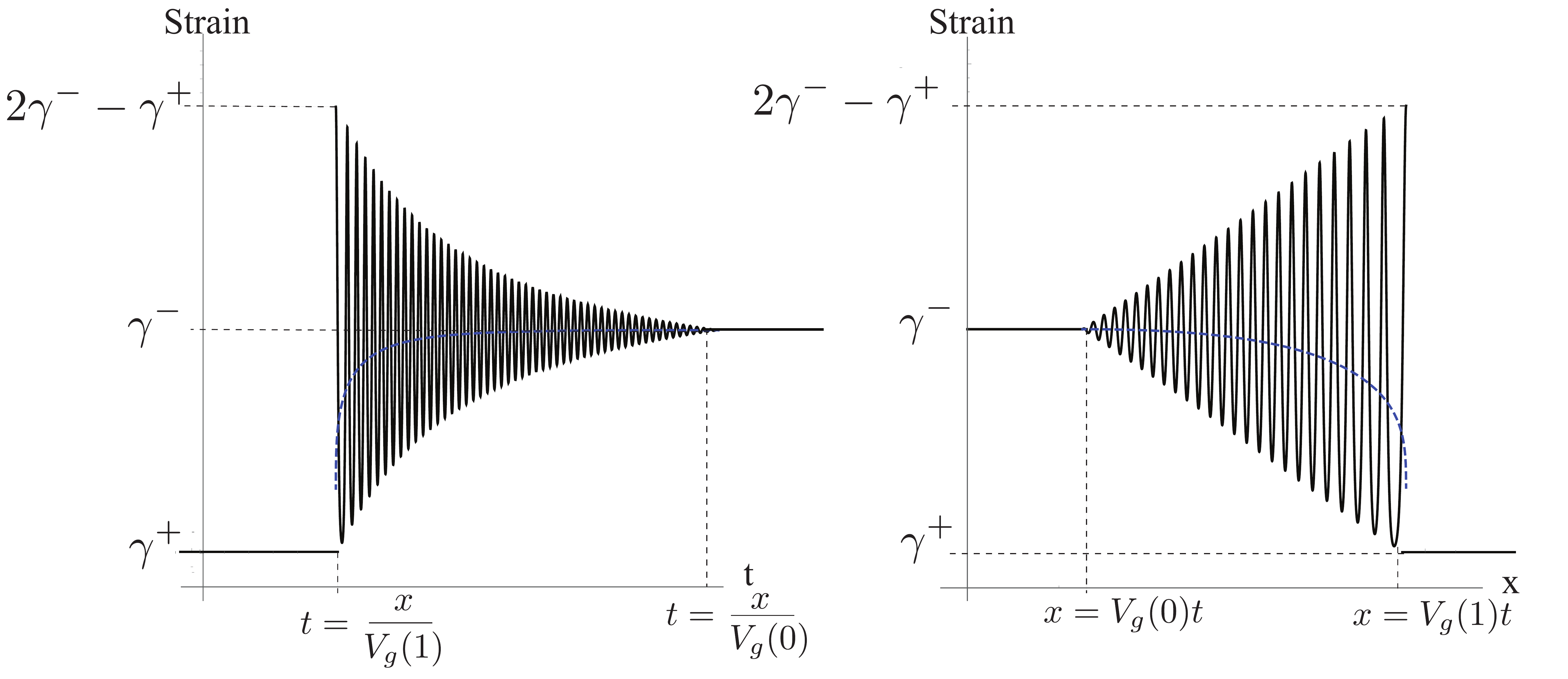}
\caption{\footnotesize Strain $\gamma(x,t)$  versus  $t$ at fixed $x$ (left) and versus $x$ at fixed  $t$ (right) according to \eqref{eq-0407-3} and \eqref{eq-20210622-1}. The leading and trailing edges travel at the respective speeds $V_g(1)$ and $V_g(0)$. The strain in front of the leading edge is $\gamma^+$, that behind the trailing edge is $\gamma^-$. The average strain given by \eqref{eq-20210520-1}, \eqref{eq-0407-3} are the dashed curves. In Problem 3 we have $\gamma^+ = 0$.}
\label{Fig-13x.pdf}
\end{center}
\end{figure}

Figure \ref{Fig-13x.pdf} shows a typical strain profile according to \eqref{eq-0407-3} and \eqref{eq-20210622-1}; the figure on the left plots $\gamma(x,t)$ versus $t$ at fixed $x$, and that on the right shows the variation of $\gamma(x,t)$ with $x$ at fixed $t$.

The approximate solution we have constructed is not appropriate at large values of the impact speed.  The trailing edge of the DSW, $x=V_g(0)t$, must lie in the first quadrant of the $x, t$-plane.
According to \eqref{eq-0406-44}, this requires $\gamma^- <  3 \beta_0^2$, which in turn by   \eqref{eq-20210621-12}, demands that
\be
\label{eq-20210621-13}
\frac{V}{c_0\beta_0^2} \ < \ 2\big[3\sqrt{3} - 2 \sqrt{2}\big] \ \approx \ 4.735. 
\ee

\begin{figure}[h]
\begin{center}
\includegraphics[scale=1.0]{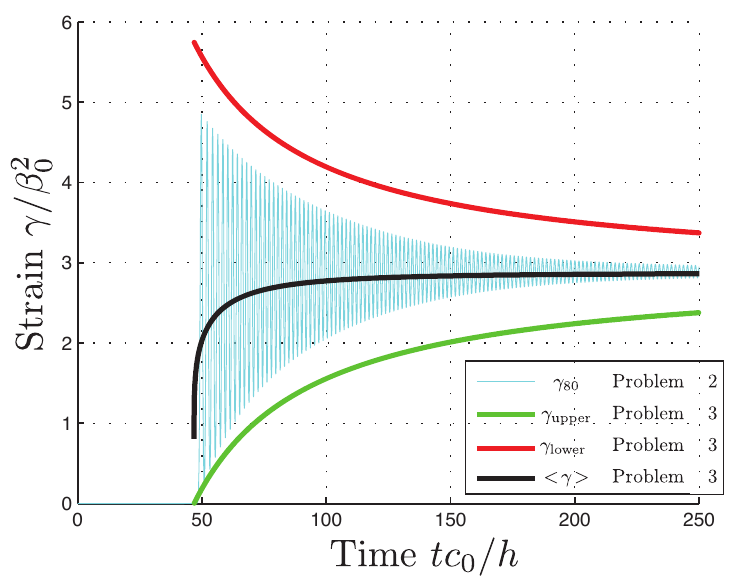}
\caption{\footnotesize Strain $\gamma_j(t)/\beta_0^2$ versus time  for spring $j=80$ (in Problem 2). The upper and lower strain envelopes $\gamma_{\rm upper}/\beta_0^2$ and $\gamma_{\rm lower}/\beta_0^2$, and the average strain $\big<\gamma\big>$  (in Problem 3) at $x=jh = 80h$ have been superposed. The figure is drawn for $V/(c_0 \beta_0^2)= 4.48, N=800$.}
\label{Fig-10.pdf}
\end{center}
\end{figure}

The equations \eqref{eq-20210612-5} governing Problem 3 were not derived from  the equations  \eqref{eq-20210511-1}, \eqref{eq-20210516-2}  governing Problem 2  (or vice versa) and the {\it detailed} solutions to the two problems do not coincide.  Even so, it is natural to 
compare some of the {\it overall features} of the solutions to the two problems. As shown already in Figure \ref{Fig-cleading.pdf},  the leading edges of the wave packets in the two problems travel at essentially the same phase speed. The trailing edge of the wave packet in Problem 2 is difficult to identify and so a similar comparison  was not attempted there. Observe from Figure \ref{Fig-13x.pdf} that the amplitude of oscillation varies curvilinearly with $t$  (left) and almost linearly with $x$ (right), similar to that in Problem 2 (Figures \ref{Fig-0.pdf} and \ref{Fig-3.pdf}).  
Since ${\rm cn}^2(z,m)$ varies between zero and one, it follows from  \eqref{eq-20210622-1} that the upper and lower envelopes of the strain field in Problem 3 are characterized by
\be
\label{eq-20210520-3}
\gamma_{\rm upper}(m) \coloneqq (1 + m^2)\gamma^- , \qquad \gamma_{\rm lower}(m) \coloneqq (1 - m^2)\gamma^-, \qquad m=m(x/t). 
\ee
Figure \ref{Fig-10.pdf}  shows a superposition of the numerical solution to Problem 2 and the envelopes $\gamma_{\rm upper}$ and $\gamma_{\rm lower}$, and the average strain $\big<\gamma\big>$, of Problem 3. 
The curvilinear (with respect to $t$) variation of the amplitude  noted previously is visible here also, and  is governed  by \eqref{eq-20210520-3}.  A similar plot versus $x$ (not shown) displays a linear variation of the amplitude in accordance with \eqref{eq-20210520-3}.

\subsection{Oscillatory energy. Apparent dissipation}\label{sec-3d}

From $\sigma_x = \rho v_t$,  \eqref{eq-20210612-3}$_1$  and  \eqref{eq-20210612-5}$_2$ one can {\it derive} the local conservation law 
\be
\label{eq-20210613-1}
\frac{\partial {\cal P}}{\partial x}= \frac{\partial {\cal E}}{\partial t}, 
\ee
where
\be
{\cal P} = \sigma v  - \eta h^2 \gamma_x \gamma_t, \qquad {\cal E} = W   - \frac 12 \eta h^2 \gamma_x^2 + \frac 12 \rho v^2.
\ee
Here $\cal P$ represents the power density (rate of working per unit length) and $\cal E$ is the {\it energy density} and so \eqref{eq-20210613-1} is simply a statement of the elastic power identity (``conservation of energy''). 
 In particular, the second term\footnote{For energetic reasons one might therefore be inclined to let $\eta$ have a negative value. However as noted previously since our goal is to mimic the discrete particle chain, we have taken $\eta$ to be positive; see first paragraph of Section \ref{sec-3}.}  in $\cal E$ can be identified with the energy associated with the strain-gradient term and the second term in $\cal P$ as the corresponding working of the associated ``couple-stress''.

According to the solution described schematically in Figure \ref{Fig-xtPlane.pdf},  the strain and particle speed at each particle
eventually settle down at the respective values $\gamma^-$ and $v^- = -V$.  Therefore at any point within or behind the DSW we set
\be
\label{eq-ss-21}
\gamma_{\rm osc}(x, t) \coloneqq \gamma(x,t) - \gamma^-, \qquad v_{\rm osc}(x, t) \coloneqq v(x,t) - v^- .
\ee
Note that $\gamma_{\rm osc}$ and $v_{\rm osc}$ vanish behind the DSW,  while within it, they represent the oscillatory parts of the strain and particle speed.  We define the energy density associated with the oscillatory part of the motion to be\footnote{Even though $\Phi(x,t) = (k(x,t) x - \omega(x,t)t)/h$ it still follows that $\partial \Phi/\partial x = k/h$ and $\partial \Phi/\partial t = - \omega/h$; see section S3 of the electronic supplemental material.}
\be
\label{eq-ss-22}
{\cal E}_{\rm osc} \coloneqq W(\gamma_{\rm osc})    - \frac 12 \eta k^2 \big(\gamma_{\rm osc}'\big)^2 + \frac 12 \rho v_{\rm osc}^2,
\ee
where $\gamma_{\rm osc}'$ is the derivative of $\gamma_{\rm osc}$ with respect to $\Phi$.
This is the oscillatory or excess part of the energy density.  The particles behind the DSW have zero oscillatory energy.

An alternative definition of the oscillatory strain and particle speed is 
\be
\label{eq-ss-21b}
\gamma_{\rm osc}(x, t) \coloneqq \gamma(x,t) - \big<\gamma\big>, \qquad v_{\rm osc}(x, t) \coloneqq v(x,t) - \big<v\big> ,
\ee
where, in the DSW, the average strain $\big<\gamma\big>$ and average particle speed $\big<v\big>$ are given by \eqref{eq-20210520-1} and \eqref{eq-20210627-1} specialized to Problem 3. The associated energy density is again given by \eqref{eq-ss-22}.

Observe that the right-hand side of \eqref{eq-ss-22} can be expressed as a function of $m$ and $\Phi$ and so we can write 
${\cal E}_{\rm osc}(x,t) = \widehat{\cal E}_{\rm osc}(\Phi(x, t), m(x, t))$. We now average this energy density over the fast oscillations to get
\be
\label{eq-ss-23}
\big<{\cal E}_{\rm osc}\big>(m) = \frac{1}{2\pi} \int_0^{2\pi} \widehat{\cal E}_{\rm osc}(\Phi, m)\, d\Phi .
\ee
Finally, integrating $\big<{\cal E}_{\rm osc}\big>(m)$ over the DSW tells us that the total oscillatory energy at time $t$ 
\be
\label{eq-20210701-3}
= \int_{V_g(0)t}^{V_g(1)t}  \big<{\cal E}_{\rm osc}\big> (m(x,t)) \, dx 
= t \int_{0}^{1} \big<{\cal E}_{\rm osc}\big>(m) V'_g(m)\, dm,
\ee
where we have used $x = V_g(m) t$ in getting the second expression.
The time {\it rate of increase of the total oscillatory energy} is therefore\footnote{We also calculated this without averaging. In this case we integrated ${\cal E}_{\rm osc} \coloneqq W(\gamma_{\rm osc})    - \frac 12 \eta h^2 \gamma^2_{x} + \frac 12 \rho v_{\rm osc}^2$ across the DSW to determine the total oscillatory energy $E_{\rm osc}(t)$. We plotted ${E}_{\rm osc}(t)$ versus $t$ where the typical graph involved oscillations about a mean straight line. The slope of this straight line provided an estimate of the rate of increase of the total oscillatory energy, ${\cal D}$. The two methods of calculation gave essentially the same results.}
\be
\label{eq-ss-24}
{\cal D} \coloneqq \int_{0}^{1} \big<{\cal E}_{\rm osc}\big>(m) V'_g(m)\, dm.
\ee

\begin{figure}[h]
\centerline{\includegraphics[scale=0.35]{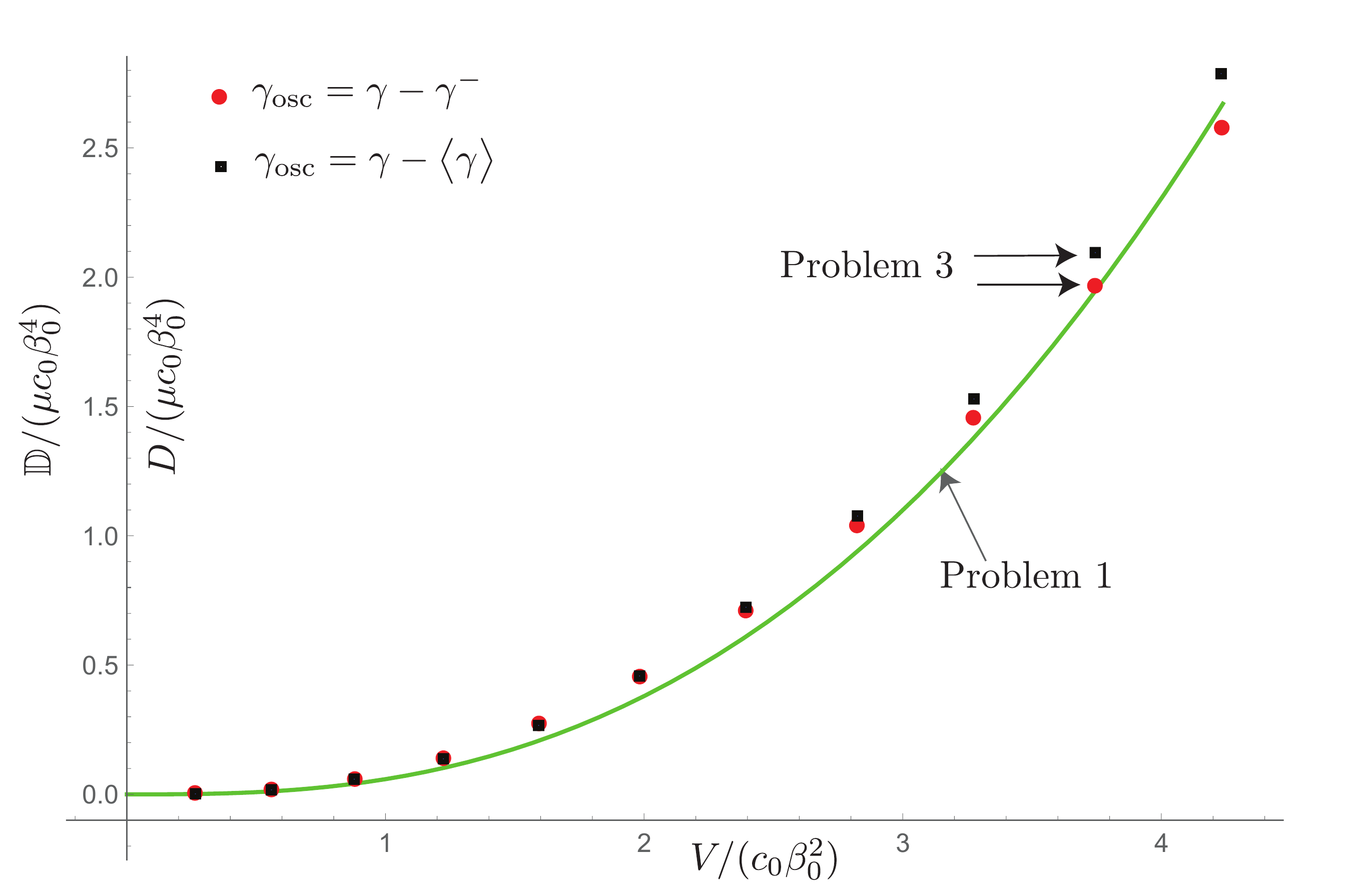}}

\caption{\footnotesize{Rate of increase of the total oscillatory energy $D/(\mu c_0 \beta_0^4)$ in Problem 3 versus impact speed $V/(c_0 \beta_0^2)$:  the dots and squares  in the figure correspond to the respective definitions \eqref{eq-ss-21}  and \eqref{eq-ss-21b}  of the oscillatory strain and speed.  When only one dot/square is visible, the other is right behind it. The dissipation rate  ${\Bbb D}/(\mu c_0 \beta_0^4)$ in Problem 3 is the solid curve. The value of the strain-gradient parameter underlying the calculations for Problem 3 was  $\eta_0 = 1/12$. We also ran calculations using $\eta_0 = 0.00001$ and could hardly tell the difference.}}

\label{Fig-DissipationRate.pdf}
\end{figure}

At each impact speed $V$, we first determined $\gamma^-, \gamma(x,t)$ and $v(x,t)$ as described in the preceding sub-section.  Then, for each definition \eqref{eq-ss-21} and \eqref{eq-ss-21b} of the oscillatory strain and particle speed, we calculated ${\cal E}_{\rm osc}$ using \eqref{eq-ss-22}; averaged it using \eqref{eq-ss-23}; and finally calculated the rate of increase of the oscillatory energy, $\cal D$, using  \eqref{eq-ss-24}. 
Such calculations were carried out for several impact speeds (consistent with \eqref{eq-20210621-13}) and the results are shown in Figure \ref{Fig-DissipationRate.pdf}.  The dots and squares in the figure correspond to the respective definitions \eqref{eq-ss-21}  and \eqref{eq-ss-21b}  of the oscillatory strain and speed. The solid curve is the dissipation rate in Problem 1 according to \eqref{eq-20210511-5}.

The average strain $\big<\gamma\big>$ is smaller than $\gamma^-$ in the interior of the DSW since $\big<\gamma\big>$ decreases monotonically from $\gamma^-$ at the trailing edge to zero at the leading edge. This presumably is why the associated rate of change of the oscillatory energy is slightly larger for \eqref{eq-ss-21b} compared to \eqref{eq-ss-21} -- the squares are above the dots.

\section{Concluding remarks.}  \label{sec-5}

This paper was focused on quantitatively understanding the  energy dissipated at a shock wave in a nonlinearly elastic bar in terms of the energy in the oscillations in two related dissipationless, dispersive systems. We considered three one-dimensional problems:  Problem 1 concerned a nonlinearly elastic bar, Problem 2 a discrete chain of particles, and Problem 3 a continuum model with a strain gradient term in the constitutive relation. Each system was semi-infinite and initially at rest in a stress-free state.  The free boundary was subjected to a sudden speed $V$ at time $t=0^+$ that was held constant from then on. Problem 2 was solved numerically, and an approximate solution to Problem 3 was constructed using modulation theory. 
For both Problems 2 and 3 we calculated the rate of increase of the oscillatory energy and compared it with  the dissipation rate at the shock in Problem 1. The results shown in Figures \ref{Fig-DissProb2.pdf} and \ref{Fig-DissipationRate.pdf} suggest that the former is a good measure of the latter. It would be interesting to prove this rigorously (if indeed it is true), most probably in the dispersionless limit of Problems 2 and 3. 

In this paper we defined the oscillatory part of the strain $\gamma_{osc}$ to be the difference between the strain and some base value of strain,  where for the base strain we considered two alternatives, $\big<\gamma\big>$ and $\gamma^-$.  The oscillatory part of the particle speed, $v_{osc}$, was defined similarly.  These seem to be fairly natural definitions.  On the other hand it is less obvious as to how to quantify the ``oscillatory energy'' (the ``energy in the oscillations'').  We considered the difference between various energies including the total energy in the DSW, the average of the total energy in the DSW, the energy associated with the average strain and speed in the DSW, the energy behind the DSW and so on. For example one alternative candidate we looked at was 
$$
\left<  W(\gamma) + \frac 12 \rho v^2 \right> - \left( W\left(\big<\gamma\big> \right)+ \frac 12 \rho \big<v\big>^2 \right).
$$ 
The dissipation rate based on the alternatives we considered did not come close\footnote{In the Supplemental Material we show the results based on two such alternative definitions.} to that in Problem 1.   The fact that the definition we eventually decided to use, $E_{osc} = W(\gamma_{osc}) + \frac 12 \rho v_{osc}^2$, gave a dissipation rate close to that in Problem 1 is not a proof that this is the correct notion of the oscillatory energy. More careful analysis of this is needed.


\noi {\bf Declaration of Competing Interests}
The authors declare that they have no known competing financial interests or personal relationships that could have appeared to influence the work reported in this paper.

\noi {\bf Acknowledgements.}  The authors gratefully acknowledge valuable feedback from Phoebus Rosakis on a first draft of this manuscript.  RA also thanks Zhantao Chen for his guidance with MATHEMATICA. PKP acknowledges partial support from a seed grant from the MRSEC at the University of Pennsylvania, grant number NSF DMR-1720530.


\bibliography{BibFile_RA_2021.bib} 


\section{Appendix.}

In Problem 3, the constitutive equation for stress, $\sigma = \mu \gamma + \frac 12 \alpha^2 \gamma^2 + \eta h^2 \gamma_{xx}$, had $\eta >0$. This was motivated by the form of the continuum equation arrived at by Taylor expanding the discrete equations, e.g. \cite{Rosenau1986}.  However this leads to instability if the wave length of a perturbation is too small (i.e. the wave number is too large).  In this section we find the condition for linear stability, and confirm that the wave numbers within the DSW conform to it. Thus, if we limit attention to perturbations whose wave numbers are close to those in the DSW, linear stability is maintained.

Consider a point within the DSW where the strain and particle speed are $\overline \gamma$ and $\overline v$.   To examine the stability of a steady uniform motion corresponding to this strain and particle speed, we substitute $\gamma = \overline \gamma + u_x, v = \overline v + u_t$
into the constitutive relation, and the result into the equation of motion. After linearization this leads to
\be
\label{eq-a0}
\mu u_{xx} + \alpha^2 \overline\gamma u_{xx} + \eta h^2 u_{xxxx} = \rho u_{tt}.
\ee
A steady periodic traveling wave solution of this linear equation has the form
\be
\label{eq-a1}
u(x,t) = {\rm exp}\,  i \! \left( \frac{kx - \omega t}{h}\right),
\ee
where $k$ and $\omega$ are constants. Keep in mind that $u$ is the perturbation and $k$ is the wave number of the perturbation. Equations \eqref{eq-a0} and \eqref{eq-a1}  lead to the dispersion relation
$$
\omega^2/c_0^2 = (1 + \overline\gamma/\beta_0^2) k^2 - \eta_0 k^4.
$$
The right-hand side of this is negative when $k$ is large, and this leads to imaginary values for $\omega$, and the corresponding perturbation \eqref{eq-a1} becomes unbounded as $t \to \infty$.  Thus linear stability requires the right-hand side of the dispersion relation to be nonnegative and so the wave number $k$  must obey
\be
\label{eq-a2}
 1 + \overline\gamma/\beta_0^2 \,  \geq \, \eta_0 k^2.
\ee

The inequality \eqref{eq-a2} is always violated if the wave number of the perturbation is sufficiently large.  However, we now show that the wave numbers within the DSW satisfy \eqref{eq-a2}. Locally, at each point within the DSW, the strain has the  mean value $\big<\gamma\big>$ and wave number $k(m)$ given by \eqref{eq-20210701-1} and \eqref{eq-0406-10} respectively.  Replacing $\overline\gamma$ and $k$ in \eqref{eq-a2} by these expressions leads to
$$
1 \geq \left[ \frac{\pi^2/6}{K^2(m)} + 1 - m^2 - 2 \frac{E(m)}{K(m)} \right] \frac{\gamma^-}{\beta_0^2}.
$$
The term within the square brackets is negative and so this inequality holds automatically for all $\gamma^- >0$.

Instead, if we replace $\overline\gamma$ by the smallest  value of the strain, $\gamma_{\rm lower}$,  given by \eqref{eq-20210520-3} (and $k$ by \eqref{eq-0406-10}), equation \eqref{eq-a2} yields
$$
1 \geq \left[ \frac{\pi^2/6}{K^2(m)} - 1 + m^2  \right] \frac{\gamma^-}{\beta_0^2}.
$$
The term in square brackets is positive and its maximum value is $\approx 0.148$ and so this inequality holds provided $\gamma^-/\beta_0^2 \lessapprox 1/(0.148) = 6.75$. Recall from the line just above \eqref{eq-20210621-13} that we  restrict attention to strains $\gamma^- < 3 \beta_0^2$.

 Thus the wave numbers in the DSW lie within the range of linear stability given by \eqref{eq-a2}.  Therefore if the wave number of a perturbation is close the wave numbers within the DSW, we have stability against such a perturbation.

\end{document}


\begin{center}
{\large \textsc{Electronic Supplementary Material}}\\
to accompany\\[2ex]

{\large \bf On the dissipation at a shock wave in an elastic bar}\\[2ex]

by\\[2ex]

Prashant K. Purohit$^{1}$ and Rohan Abeyaratne$^{2}$\\[2ex]

$^{1}$Department of Mechanical Engineering and Applied Mechanics, University of Pennsylvania,  Philadelphia, Pennsylvania, 19104, USA\\
$^{2}$Department of Mechanical Engineering, Massachusetts Institute of Technology, Cambridge,  Massachusetts, 02139, USA

July 20, 2021
\end{center}


\renewcommand{\theequation}{\arabic{equation}}
\setcounter{equation}{0}

\noi\hrulefill

\section{Supplemental material.}


\renewcommand{\theequation}{\arabic{equation}}
\setcounter{equation}{0}

\noindent {\large {\bf S1.  Numerical schemes used in Problem 2 conserved energy.}}

\begin{figure}[h]
\begin{center}
\includegraphics[scale=0.8]{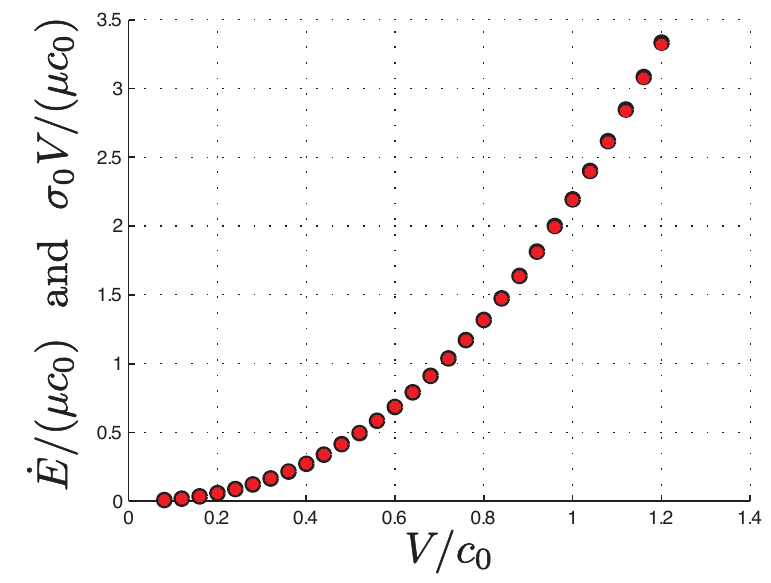}
\caption{\footnotesize The rate of change of the total energy $\dot E/(\mu c_0)$ (red) and the rate of working $\sigma_0 V/(\mu c_0)$ (blue) versus the impact speed $V/c_0$ in Problem 2.}
\label{Fig-Econs}
\end{center}
\end{figure}

We verified that the numerical methods used in Problem 2 conserved energy and that there was no numerical dissipation:  we first calculated and plotted the total (kinetic plus potential) energy $E(t)$ versus $t$ for several impact speeds and observed that the relationship was linear (with superposed oscillations of extremely small amplitude). We were thus able to unambiguously identify $\dot E = dE/dt$ and plot  it versus $V$; see the red  filled  circles in Figure \ref{Fig-Econs}.  The figure also shows the rate of working $\sigma_0 V$ (blue open circles)  versus the impact speed $V$, where the force in the zeroth spring $\sigma_0 = U'(\overline\delta)$ was determined from the numerical solution. 
The coalescence of these points in Figure \ref{Fig-Econs} confirms that the numerical scheme conserves energy.


\noi\hrulefill
\renewcommand{\theequation}{\arabic{equation}}
\setcounter{equation}{0}

\noindent {\large {\bf S2.  Showing $E_{osc}(t)$ vs. t in Problem 2}}

\begin{figure}[h]
\begin{center}
\includegraphics[scale=0.8]{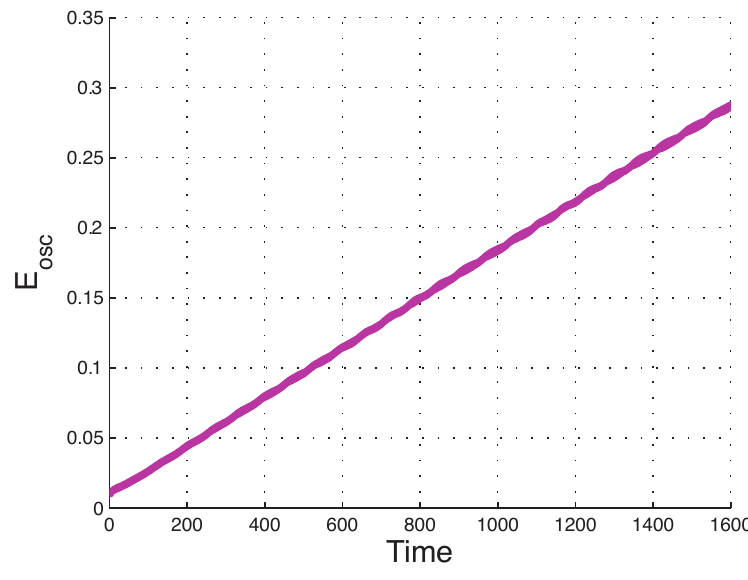}
\caption{\footnotesize $E_{ocs}(t)$ versus $t$ in Problem 2 for $\frac{V}{c_{0}\beta_{0}^{2}} = 4.48$ and $N = 800$.}
\label{Fig-Supl-1}
\end{center}
\end{figure}

At a given time $t$ we first determine the leading edge of the wave in Problem 2 by finding the spring $n(t)$ at which $\gamma_{n} \geq 1.0\times 10^{-8}$. We verified that changing the tolerance to 
$1.0 \times 10^{-6}$ in this criterion does not change the position of the leading edge. It is assumed that
$\gamma_{j} = 0$ and $v_{j} = 0$  for $j > n$. Next, for $j \leq n$ we compute 
$\gamma^{osc}_{j}(t) = \gamma_{j}(t) - \gamma^{-}$ and $v^{osc}_{j}(t) = v_{j} - v^- = v_{j} + V$. Then, $E_{osc}(t) = \sum_{j=0}^{n(t)} \frac{m}{2}(v^{osc}_{j}(t))^{2} + W(\gamma^{osc}_{j}(t))$. A plot of 
$E_{osc}(t)$ versus $t$ for ${V}/{(c_{0}\beta_{0}^{2})} = 4.48$ is shown in Figure \ref{Fig-Supl-1}. The plot is a straight line with superimposed oscillations of low amplitude and high frequency. The
slope of the line gives the rate of increase of the oscillatory energy  in Problem 2 that we identify with the dissipation rate in Problem 1.   


\noi\hrulefill
\renewcommand{\theequation}{\arabic{equation}}
\setcounter{equation}{0}

\noindent {\large {\bf S3.  Showing that $\Phi_x =  k/h, \Phi_t = - \omega/h$}}

When $k$ and $\omega$ are constant, it is obvious that $\Phi_x =  k/h, \Phi_t = - \omega/h$ where $\Phi = (kx - \omega t)/h$. To show that this continues to be true for our choice of $\Phi(x,t) = (k(x,t) x - \omega(x,t) t)/h$ we proceed as follows using the fact that $k$ and $\omega$ depend on $x,t$ only though $m(x,t)$:
\be
\label{eq-sm-1}
\Phi(x,t) = \frac{k(m) x - \omega(m)t}{h}, \qquad x/t = V_g(m), \qquad m = m(x/t).
\ee
To establish this we proceed as follows:
\be
\ba{lll}
\displaystyle \frac{\partial \Phi}{\partial x} &\displaystyle = \frac kh + \frac 1h \left( x k'(m) \frac{\partial m}{\partial x} - t \, \omega'(m) \frac{\partial m}{\partial x} \right) =   \frac kh + \frac 1h \Big( x   - t \, \omega'(m)/k'(m) \Big) k'(m) \frac{\partial m}{\partial x}  =\\[2ex]

&\displaystyle =   \frac kh + \frac 1h \left( x   - t \, \frac{d\omega}{dk} \right) k'(m) \frac{\partial m}{\partial x} 
= \displaystyle =  \frac kh + \frac 1h \left( x   - V_g t\right) k'(m) \frac{\partial m}{\partial x} \stackrel{\eqref{eq-sm-1}_2}{=} \frac kh
\ea
\ee
An analogous calculation establishes $\Phi_t = -\omega/h$.


\noi\hrulefill

\renewcommand{\theequation}{\arabic{equation}}
\setcounter{equation}{0}

\noindent {\large {\bf S4.  Converting equations governing a 1-D elastic bar from Lagrangian to Eulerian form. }}

The motion is described by $y=y(x, t)$. If we have a function $f(x, t)$ we write its $x$ and $t$ partial derivatives as $f_x$ and $\dot f$;  if we have a function $f(y,t)$ we write its $y$ amd $t$ partial derivatives as $f_y$ and $f_t$. By the chain rule we have the relation
\be
\label{eq1}
\dot f = f_y v + f_t \qquad {\rm where} \qquad v = \dot y.
\ee

The basic Lagrangian equations for the bar are
\be
\label{eq2}
\lambda = y_x, \qquad v = \dot y, \qquad v_x = \dot \lambda, \qquad \sigma = W'(\lambda), \qquad \sigma_x = \rho_0 \dot v,
\ee
where we now use $\rho_0$ to denote the mass density in the reference configuration. Here $\lambda = 1 + \gamma$ is the stretch.
{\it First we derive the counterpart of $v_x = \dot\lambda$}: Introduce the variable $\rho$, defined in terms of $\lambda$, by
\be
\label{eq3}
\rho = \rho_0/\lambda, \qquad \rho\lambda = \rho_0.
\ee
It is immediately apparent that  $\rho$ is the mass density in the current configuration and \eqref{eq3} is a statement of mass balance.
Now
\be
\label{eq4}
\eqref{eq3}_2 \quad \Rightarrow \quad \dot \rho \lambda + \rho \dot \lambda = 0 \quad \stackrel{\eqref{eq1}}{\Rightarrow} \quad (\rho_y v +\rho_t)\lambda +  \rho \dot \lambda = 0
\stackrel{\eqref{eq2}_3}{\Rightarrow} \quad (\rho_y v +\rho_t)\lambda +  \rho v_x = 0.
\ee
By the chain rule $v_x = v_y y_x = v_y\lambda$. Therefore
\be
\label{eq5}
\eqref{eq4} \quad \Rightarrow \quad  (\rho_y v +\rho_t)\lambda +  \rho v_y \lambda = 0 \quad \Rightarrow \quad \rho_t + (\rho v)_y = 0.
\ee
This is the mass balance equation in the form customarily written in fluid mechanics. 

\noi {\it Second transform the constitutive equation:}  Define the energy per unit current length $U(\rho)$ in terms of the energy per unit reference length $W(\lambda)$ by
\be
\label{eq6}
W(\lambda) = \rho_0 U(\rho).
\ee
Differentiate both side by $\lambda$ and use \eqref{eq2}$_4$
\be
\sigma = W'(\lambda) = \rho_0 U'(\rho) \frac{d\rho}{d\lambda} \stackrel{\eqref{eq3}_1}{=} - \frac{\rho_0^2}{\lambda^2} U'(\rho) \stackrel{\eqref{eq3}_1}{=} - \rho^2 U'(\rho).
\ee
Therefore the constitutive relation for the pressure $-\sigma$ is
\be
\label{eq7}
- \sigma = \rho^2 U'(\rho).
\ee
This is the form of the constitutive equation as customarily written in fluid mechanics.  One of the commonly used pressure-density relations in fluids is $-\sigma = c \rho^n$ (ideal gas undergoing a polytropic process) which maps into $\sigma = - c \rho_0^n/\lambda^n$.

\noi {\it Third we derive the counterpart of $\sigma_x = \rho_0 \dot v$}:  By the chain rule $\sigma_x = \sigma_y y_x = \sigma_y \lambda$. So
\be
\eqref{eq2}_5 \quad \stackrel{\eqref{eq1}}{\Rightarrow} \quad \sigma_y \lambda = \rho_0 (v_yv + v_t) \quad
\stackrel{\eqref{eq3}}{\Rightarrow} \quad   \sigma_y  = \rho (v_y v + v_t) \quad \Rightarrow\quad
 \sigma_y  = \rho vv_y  + (\rho v)_t - \rho_t v.
\ee
Now use  \eqref{eq5} in here
\be
 \sigma_y  = \rho vv_y  + (\rho v)_t + v (\rho_y v + \rho v_y) = (\rho v^2)_y + (\rho v)_t ,
\ee
and finally use \eqref{eq7}
\be
\label{eq8}
\frac{\partial}{\partial y}\left( \rho^2 U'(\rho) + \rho v^2\right) + \frac{\partial}{\partial t}(\rho v) = 0.
\ee
This is the equation of motion in the form customarily written in fluid mechanics. 

Note that equations \eqref{eq5} and \eqref{eq8} each involves both $\rho$ and $v$.


\noi\hrulefill

\noindent {\large {\bf S5.  Four conservation laws for deriving Whitham's modulation equations. }}

\renewcommand{\theequation}{\arabic{equation}}
\setcounter{equation}{0}

The steady periodic traveling waves we found involved four parameters $\gamma^+, \gamma^-, m$ and $v_*$.  The most general modulated waves constructed from them would allow these parameters to be four slowly varying functions that can be determined by averaging four conservation laws.  One choice of such a set of conservation laws for the setting considered in this paper is   
$$
\frac{\partial}{\partial x} \left(v\right) =  \frac{\partial}{\partial t}\left( \gamma \right),
$$
$$
\frac{\partial}{\partial x} \left(\sigma\right) = \rho \frac{\partial}{\partial t}\left( v\right),
$$
$$
\frac{\partial}{\partial x} \left(\sigma v  - \veps^2 \gamma_xv_x \right) = \frac{\partial}{\partial t}\left( W - \frac 12 \veps^2 \gamma_x^2 + \frac 12 \rho v^2 \right),
$$
$$
\frac{\partial}{\partial x} \left(W(\gamma)  + \frac 12 \veps^2 \gamma^2_{x}  - \sigma \gamma - \frac 12 \rho v^2 \right) = \frac{\partial}{\partial t}\left( -\rho \gamma v \right),
$$
where $\sigma = W'(\gamma) + \veps^2 \gamma_{xx}$. The first of these describes kinematic compatibility and the second linear momentum balance.  The third conservation law is obtained by multiplying the second by $v$ (and carrying out some manipulations), while the fourth is likewise obtained by multiplying the second by $\gamma$ (and carrying out some manipulations). The third equation represents the elastic power identity (conservation of energy.) The fourth equation appears to be the one-dimensional version of the conservation law obeyed by Eshelby's energy-momentum tensor (e.g. see Section 5.4.5 of [1]) augmented with the strain-gradient term.

\noi 1. R. W. Ogden, Non-linear Elastic Deformations, Dover, 1997.

\noi\hrulefill


\noindent {\large {\bf S6. Alternative definitions of the oscillatory energy. }}

\renewcommand{\theequation}{\arabic{equation}}
\setcounter{equation}{0}

As mentioned in the concluding remarks in Section 5, we considered several definitions of the oscillatory energy. Here we present two other definitions we used and the corresponding results.

Let definition 1 be the one we used in the paper.  In definition 2 we calculated $E_{osc}$ using
\begin{equation}
E_{\rm osc}(t) \coloneqq \int_{V_{g}(0)t}^{V_{g}(1)t}\left[\left(W(\gamma) + \frac{\rho}{2}v^{2}\right) - \left(W(\langle \gamma \rangle) + \frac{\rho}{2}\langle v\rangle^{2}\right)\right]\,dx,
\end{equation}
where $\gamma(x,t)$ and $v(x,t)$ are the strain and particle speed in the DSW as given by equations (59) and (60) specialized to Problem 3, and $\big< \gamma \big>$ and $\big< v \big>$ are the average strain and particle speed. 
In definition 3 we set  
\begin{equation}
E_{\rm osc}(t) \coloneqq  \int_{V_{g}(0)t}^{V_{g}(1)t}\left[\left(W(\gamma) + \frac{\rho}{2}v^{2}\right) - \left(W(\gamma^{-}) + \frac{\rho}{2}(v^{-})^{2}\right)\right]\,dx,
\end{equation}
where $\gamma^-, v^-$ are the strain and particle speed behind the DSW. 

\begin{figure}[h]
\begin{center}
\includegraphics[scale=0.8]{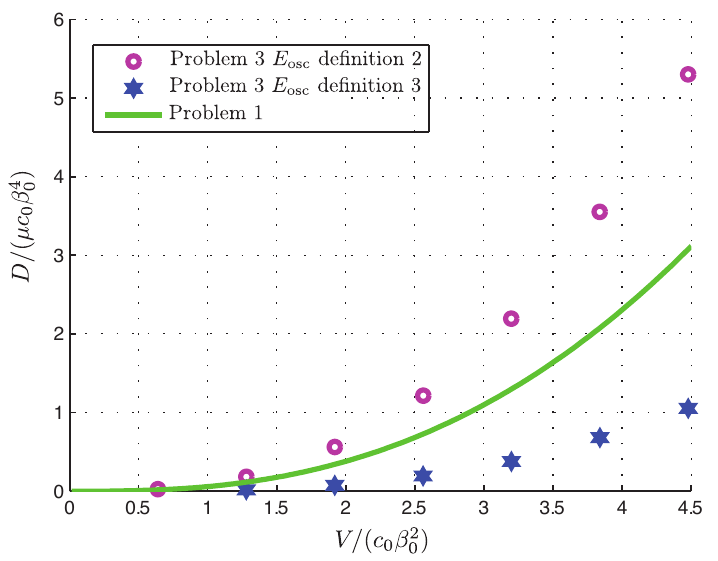}
\caption{\footnotesize Non-dimensional dissipation rate versus impact speed ${V}/(c_{0}\beta_{0}^{2})$ in Problem 3 computed using two alternative definitions of $E_{\rm osc}$. The solid green curve corresponds to Problem 1.}
\label{Fig-Supl-2}
\end{center}
\end{figure}

The dissipation rate corresponding to each definition, calculated using ${\cal D} = {d E_{\rm osc}}/{dt}$, is shown in Figure \ref{Fig-Supl-2}. Definition 2 (magenta circles) overestimates the dissipation rate compared to Problem 1  and definition 3  (blue stars) underestimates it. 

